\documentclass[aps,prf,superscriptaddress]{revtex4-2}

\usepackage{dcolumn}
\usepackage{bm}
\usepackage{caption}
\usepackage{booktabs}
\usepackage{soul}
\sethlcolor{yellow}
\usepackage{xcolor}
\usepackage{graphicx}
\usepackage{xcolor}
\usepackage{newtxtext}
\usepackage{newtxmath}
\usepackage{natbib}
\usepackage{subcaption}
\usepackage{hyperref}
\hypersetup{
    colorlinks=true,      
  linkcolor=blue,       
  citecolor=blue,       
  urlcolor=blue         
}

\newcommand{\RomanNumeralCaps}[1]
\linenumbers

\begin{document}

\preprint{APS/123-QED}

\title{Electric modification of mode competition in viscous films with insoluble surfactants on vertical fibers}

\author{Jun Gao}
\affiliation{School of Astronautics, Beihang University, Beijing 100191, PR China}
\affiliation{Aircraft and Propulsion Laboratory, Ningbo Institute of Technology, Beihang University, Ningbo 315800, PR China}

\author{Xiaocong Yang}
\affiliation{School of Astronautics, Beihang University, Beijing 100191, PR China}

\author{Senlin Zhu}
\email{Author to whom correspondence should be addressed: slzhu@pku.edu.cn}
\affiliation{School of Astronautics, Beihang University, Beijing 100191, PR China}

\author{Bingqiang Ji}
\email{Author to whom correspondence should be addressed: bingqiangji@buaa.edu.cn}
\affiliation{School of Astronautics, Beihang University, Beijing 100191, PR China}

\author{Qingfei Fu}
\email{Author to whom correspondence should be addressed: fuqingfei@buaa.edu.cn}
\affiliation{School of Astronautics, Beihang University, Beijing 100191, PR China}
\affiliation{Aircraft and Propulsion Laboratory, Ningbo Institute of Technology, Beihang University, Ningbo 315800, PR China}
 
\begin{abstract}
This study investigates the coupled effects of an insoluble surfactant and a radial electric field on the stability of a viscous liquid film flowing down a vertical fiber. Starting from the governing equations in two dimensions, a reduced model in one dimension is derived using the long wave approximation to describe the coupled evolution of the interface and surfactant transport. Linear stability analysis identifies two distinct unstable modes: the Rayleigh-Plateau mode, which dominates at lower values of the Marangoni number $Ma$, and the Marangoni mode, which becomes dominant at higher values of $Ma$. The influence of the radial electric field is determined by the position of the outer electrode $\beta$. When $\beta<\mathrm{e}$, the electric field enhances both instabilities and narrows the stable interval in $Ma$ between the two modes. When $\beta>\mathrm{e}$, the electric field suppresses both modes and can completely eliminate the unstable region associated with the Marangoni mode even at a relatively small electric Weber number $E_b$. Continuation of the traveling wave solutions further shows that, when $\beta<\mathrm{e}$, the magnitude of the relative interfacial motion $I_{RP}$, generally increases with $E_b$. By contrast, the intensity of Marangoni convection $I_{M}$ varies only weakly at smaller values of $E_b$ and increases appreciably only when the electric field becomes sufficiently strong. Analysis of the stream function and the relative interfacial velocity reveals that the electric stress primarily intensifies the recirculation beneath the wave crest and reshapes the spatial distribution of the relative interfacial velocity. This redistribution modifies the advective transport of surfactant, reorganizes the interfacial concentration field, and thereby alters the Marangoni stress. The essential mechanism through which the surfactant affects film stability is therefore the competition between Marangoni convection and relative interfacial motion, whereas the electric field modifies this competition indirectly by changing the magnitude and spatial distribution of the relative interfacial velocity.
\end{abstract}

\maketitle

\section{Introduction}
\label{Introduction}
The study of falling liquid films can be traced back to the classical theory of film condensation proposed by \citet{Nusselt1916}, in which a liquid film descending along a solid wall was treated as a uniform laminar layer, while fluctuations of the free surface were neglected. This treatment yielded the fundamental solution for falling films and established the theoretical basis for subsequent investigations. Kapitza and Kapitza \citep{Kapitza1948a,Kapitza1948b,Kapitza1949} later combined theoretical analysis with experimental observations to reveal how free surface waves emerge and evolve dynamically in liquid films flowing downward along a wall. To clarify the physical origin of film instability, \citet{Benjamin1957} conducted the classical linear stability analysis of a liquid film flowing down an inclined plane and demonstrated that the system is susceptible to long wave instability. The analysis showed that a critical Reynolds number exists, $Re_c=\frac{5}{4}\cot\theta$, where $\theta$ denotes the inclination angle of the wall, and that long wave disturbances become unstable once $Re$ exceeds this critical value. \citet{Yih1963} subsequently extended the analysis by incorporating surface tension and showed that surface tension suppresses the growth of disturbances with short wavelengths. Although linear stability theory provides an accurate description of the initial amplification of infinitesimal disturbances, its applicability is largely restricted to the early stage of instability, which makes it unsuitable for describing finite amplitude waves and the nonlinear dynamics that arise after prolonged evolution. To overcome this limitation, \citet{Benney1966} derived a reduced model for the temporal and spatial evolution of the local film thickness by using the long wave approximation together with lubrication theory, thereby extending the study of falling films from linear instability to nonlinear evolution.

In contrast to falling films on inclined planes, which are usually formulated in Cartesian coordinates, films descending along fibers must be described within a cylindrical geometry. Because this geometry introduces azimuthal curvature in addition to axial curvature, the Rayleigh–Plateau (RP) instability becomes another important mechanism governing the evolution of the interface. Building on the relatively mature theory developed for inclined planes, \citet{Goren1962} performed an early linear stability analysis of an annular liquid film coating a slender fiber and showed that, for prescribed values of the radius ratio and the Ohnesorge number, a disturbance with a maximum growth rate exists. The corresponding wavelength determines the characteristic spacing between droplets after the film ruptures. \citet{Quere1990} later carried out systematic experiments on liquid films flowing down the exterior of vertical fibers, from which empirical relations connecting the film thickness, the fiber radius, and the capillary length were obtained. These relations were further used to determine whether a liquid film coating a fiber would remain continuous or evolve into a sequence of droplets. Under the assumption that the film thickness is much smaller than the fiber radius, \citet{Frenkel1992} derived the first long wave reduced model in cylindrical coordinates in which the local film thickness serves as the primary dependent variable. \citet{Chang1999} subsequently solved this model numerically and clarified the mechanism through which secondary waves generated downstream collide and merge, eventually producing large droplets. As emphasized by \citet{Kalliadasis1994}, the interfacial dynamics of a film flowing down an inclined plane are governed mainly by axial curvature, whereas a cylindrical geometry introduces azimuthal curvature, which competes with the axial contribution. The relative magnitude of these two curvature effects gives rise to distinct dynamics in the thin film and thick film regimes. It should also be noted that a liquid film descending along the interior wall of a tube is not normally classified as a thick film. Nevertheless, the internal coating problem has attracted extensive theoretical and numerical attention in recent years, as illustrated by the studies of \citet{Camassa2014,Camassa2016}.

For thick films flowing along the exterior of fibers, \citet{Kliakhandler2001} performed pioneering experiments to investigate the associated unstable dynamics. By observing castor oil descending along a nylon fiber at three different flow rates, three characteristic flow regimes were identified. Regime $(a)$ corresponds to a convective state, in which two neighbouring large droplets are separated by a long and nearly flat film. Regime $(b)$ corresponds to a Rayleigh–Plateau state, in which droplets descend in a nearly periodic train and form a structure resembling a necklace. Regime $(c)$ corresponds to a solitary wave state, in which a single large droplet is followed by a sequence of small capillary ripples.  \citet{Craster2006} subsequently introduced traveling wave solutions into the analysis and determined the wave speeds and wavenumbers associated with the three regimes. Except for regime $(a)$, however, considerable discrepancies remained between the theoretical or numerical predictions and the experimental measurements. \citet{Duprat2007} extended the experiments of \citet{Kliakhandler2001} by examining the transition between absolute and convective instability. The resulting analysis established the relationship between the flow rate and the fiber radius at the transition boundary and further demonstrated that all the experimental conditions considered by \citet{Kliakhandler2001} were located within the absolutely unstable region. \citet{RuyerQuil2008} later derived a second order reduced model for thick liquid films coating fibers by using a weighted residual formulation. Because inertia and viscous dissipation were retained more consistently, the agreement between traveling wave solutions and experimental measurements was substantially improved. Nevertheless, the capillary ripples observed behind the large droplet in regime $(c)$ could not be reproduced accurately. This discrepancy was eventually explained by \citet{Ding2019} through a theory based on relative periodic solutions. The underlying mechanism is associated with persistent nonlinear interactions between neighbouring droplets of different sizes, which cause an otherwise steadily propagating traveling wave to undergo periodic oscillations. The resulting oscillatory state can therefore be represented by a relative periodic orbit.

Radial electric fields provide an effective way to manipulate the morphology of liquid films on fibers. \citet{Li2009} investigated the axial and azimuthal instabilities of a thin film coating a fiber within a perfectly conducting fluid model, in which surface tension, van der Waals interactions, and Maxwell stresses act simultaneously. The results showed that increasing the applied electric field enhances the growth rate of disturbances while reducing the characteristic wavelength, thereby enabling control over the length scale of microscale and nanoscale structures on cylindrical surfaces. \citet{Kishore2012} further examined the competition between azimuthal curvature and electrostatic stresses, and found that bead-like structures form when the curvature effect is dominant, outward-projecting ridges arise when the electric field dominates, and mixed morphologies containing both beads and ridges appear when the two effects have comparable strengths. The studies discussed above mainly focused on thin liquid films subjected to electric fields, \citet{Wray2012,Wray2013} were among the first to extend the problem of radial electric fields to thick films coating the exterior of fibers. Using the leaky dielectric model and the long wave approximation, \citet{Wray2013} derived a reduced model for the coupled evolution of the interface and the interfacial charge. Their analysis showed that the influence of the electric field on film stability is determined jointly by the conductivity ratio, the permittivity ratio, and the position of the outer electrode. The model also accounts for both the normal and tangential components of the Maxwell stress and therefore provides a general description of the effects of finite charge relaxation. However, the analysis was conducted over a broad range of electrical parameters and was not developed specifically for the castor oil-air experiments of \citet{Kliakhandler2001}. Because the extremely low conductivity of air gives this system a very large conductivity ratio, its stability characteristics are expected to approach the limit of a perfectly conducting liquid, whereas the effects associated with interfacial charge accumulation and tangential electrical stresses become less pronounced. Motivated by the thick film experiments of \citet{Kliakhandler2001}, \citet{Ding2014} subsequently adopted the limiting assumption of a perfectly conducting liquid and examined the influence of a radial electric field on both the linear stability and nonlinear evolution of the film. Their results showed that the effect of the electric field is controlled by the outer electrode position $\beta$. The field suppresses the instability when $\beta>\mathrm{e}$ but enhances disturbance growth when $\beta<\mathrm{e}$. Analysis of traveling wave solutions further showed that the destabilizing mechanism arises because the electric field strengthens the internal circulation within the droplets. The nonlinear simulations nevertheless indicated that stabilization by the electric field is not robust over the accessible parameter range: a relatively weak field may still generate oscillatory traveling waves or even singular behaviour in which the interface touches the outer electrode, whereas pronounced suppression generally requires a high electric potential, which limits the economic feasibility of this approach in engineering applications.

Surfactants can substantially modify free-surface flows because a non-uniform interfacial distribution generates surface-tension gradients and the associated Marangoni stresses. Depending on the base flow and the interfacial transport process, surfactants may either suppress film instability or introduce additional destabilizing mechanisms. For films on planar or inclined substrates, Marangoni stresses not only modify the classical interfacial mode but may also generate an additional surfactant mode. For example, \citet{Halpern2003} showed that insoluble surfactants can destabilize creeping two-layer flow, whereas \citet{Blyth2004} found that the influence of surfactants on inclined-film flow depends sensitively on both the Reynolds number and the Marangoni number. In core-annular flows, \citet{Wei2005} and \citet{Blyth2006} further demonstrated that the coupling between the base flow and surfactant transport can produce non-monotonic variations in stability. For cylindrical geometries, \citet{Nair2020} developed a two-dimensional model for a surfactant-laden liquid film flowing over a cylindrical rod and identified both interfacial and surfactant modes through linear stability analysis. Building on the thick-film experiments of \citet{Kliakhandler2001}, \citet{Gao2026} investigated the influence of insoluble surfactants on the stability of thick films flowing down fibers and proposed a competition mechanism between the relative motion at the interface and Marangoni convection. Although an absolutely stable regime was found at intermediate Marangoni numbers, the corresponding stability window was narrow, which restricts the range of surfactants and operating conditions available for stabilization.

The limitations of electric fields and surfactants when applied separately raise two closely related questions: whether their combined action can provide a more effective means of regulating film instability, and how electric forcing modifies the competition between relative interfacial motion and Marangoni convection. To address these questions, we consider the axisymmetric flow of a perfectly conducting Newtonian liquid film down a vertical metallic fiber in the presence of an insoluble surfactant and a radial electric field. Section~\ref{Problem formulation} presents the governing equations in two dimensions, together with the interfacial stress balances and the transport equation for the surfactant. Section~\ref{Scaling and reduced model} introduces the characteristic scales and derives a reduced model in one dimension using the long wave approximation. Section~\ref{Linear stability and mode competition} examines the linear stability of the uniform film, with particular attention to the competition and transition between the Rayleigh-Plateau and Marangoni modes, the variation of the neutral stability boundaries, and the departure of the coupled response from linear superposition. Section~\ref{Nonlinear dynamics and coupling mechanisms} analyses the traveling wave solutions and clarifies how electric forcing modifies the competition between Marangoni convection and relative interfacial motion. The principal conclusions are summarised in Section~\ref{sec:conclusion}.
\begin{figure}
\centerline{\includegraphics[width=0.3\linewidth]{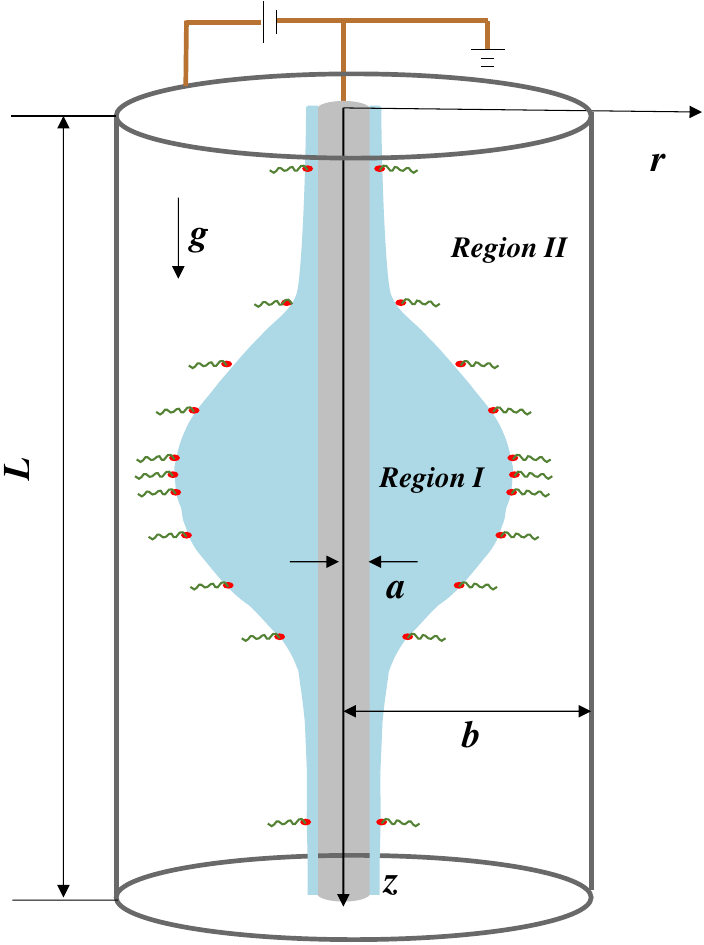}}
  \caption{Electrohydrodynamic flow of a surfactant-laden Newtonian film coating a vertical fiber.}
  \label{fig1}
\end{figure}
\section{Problem formulation.}
\label{Problem formulation}
In this study, we consider the axisymmetric gravity-driven flow of a surfactant-laden Newtonian liquid film coating a vertical cylindrical metal fiber under the influence of an externally imposed electric field, as illustrated in \autoref{fig1}. The model is formulated in cylindrical coordinates $(r^*,z^*)$, where $r^*$ and $z^*$ denote the radial and axial coordinates, respectively. Following the idealized electrohydrodynamic framework adopted by Ding et al.\cite{Ding2014}, the liquid film is treated as perfectly conducting. Under this approximation, the metal fiber and the liquid film form an equipotential region, so that the electric field is confined to the annular gas region between the liquid-gas interface and the outer electrode. The physical domain is divided into two regions. Region I consists of the metal fiber and the conducting liquid film, which share a uniform electric potential. Region II is the annular space between the liquid-film surface and the outer electrode, and is occupied by a dielectric gas whose hydrodynamic contribution is neglected. An insoluble surfactant is distributed along the liquid-gas interface, and its interfacial concentration is denoted by $\Gamma^*(z^*,t^*)$. The surfactant modifies the local surface tension through an equation of state $\gamma^*=\gamma^*(\Gamma^*)$, thereby introducing Marangoni stresses into the interfacial stress balance. Under the electrostatic approximation, magnetic induction effects are neglected, and the electric field in Region II is irrotational. Consequently, the electric potential in Region II satisfies Laplace's equation:
\begin{equation}
\label{eq:laplace-potential}
    \phi^*_{rr}+(r^*)^{-1}\phi^*_r+\phi^*_{zz}=0,
\end{equation}
where $\phi^*$ is the electric potential. Meanwhile, a high electric potential $\phi_0^*$ is applied at the outer electrode at $r^*=b^*$, while the fiber is grounded, resulting in the electric potential being zero at the liquid film surface $r^*=a^*+h^*(z^*,t^*)$ to ensure no generation of a tangential electric field. Then the Maxwell stress $\mathbf{T}=\epsilon[\mathbf{E}\mathbf{E}-\frac{1}{2}|\mathbf{E}|^2\mathbf{I}]$ can be obtained by solving the Laplace equation, where $\varepsilon_0$ is the electrical permittivity of the gas, $\mathbf{E}=-\nabla \phi^*$ is the electric intensity, and $\mathbf{I}$ is the identity tensor.

Subsequently, we begin to establish the governing equations for the liquid film flow in region I. The fluid is assumed to be Newtonian, incompressible, and perfectly conductive, with axisymmetric motion. The continuity equation and momentum equation in cylindrical coordinates are given as follows:
\begin{equation}
\label{eq:continuity}
    u_r^*+(r^*)^{-1}u^*+w_z^*=0,
\end{equation}
\begin{equation}
\label{eq:axial momentum}
\rho^*\left(u_t^*+u^*u_r^*+w^*u_z^*\right)
=-p_r^*+\mu^*\left[ u_{rr}^*+(r^*)^{-1}u_r^*+u_{zz}^*-(r^*)^{-2}u^*\right],
\end{equation}
\begin{equation}
\label{eq:radial momentum}
\rho^*\left(w_t^*+u^*w_r^*+w^*w_z^*\right)
=-p_z^*+\mu^*\left[w_{rr}^*+(r^*)^{-1}w_r^*+w_{zz}^*
    \right]+\rho^*g^*,
\end{equation}
where $u^*$ and $w^*$ stand for the radial and axial velocity of the film, respectively. $\rho^*$ is the density of the fluid, while $\mu^*$ is the dynamic viscosity, and $g^*$ is the gravitational acceleration. The no-slip and no-penetration boundary condition is satisfied at the surface of the fiber:
\begin{equation}
\label{boundary:fiber}
    u^*(a^*)=w^*(a^*)=0.
\end{equation}
At the gas-liquid interface, where $r^*=a^*+h^*(z^*,t^*)$, the kinematic boundary condition and the normal as well as the tangential stress balance must be satisfied:
\begin{equation}
\label{boundary:kinematic}
    u^*=h_t^*+w^*h_z^*,
\end{equation}
\begin{equation}
\label{boundary:normal stress}
(\Delta_h^*)^2\left(p^*-p_g^*\right)-2\mu^*\left[
u_r^*-h_z^*\left(w_r^*+u_z^*\right)+(h_z^*)^2w_z^*\right] 
 +\epsilon\left[ \frac{1}{2}\left(1-(h_z^*)^2\right)\left((\phi_r^*)^2-(\phi_z^*)^2\right)
-2h_z^*\phi_r^*\phi_z^* \right]
=2(\Delta_h^*)^2\gamma^*\vartheta^*,
\end{equation}
\begin{equation}
\label{boundary:tangential stress}
2\mu^*h_z^*\left(u_r^*-w_z^*\right)+ 
\left[2-(\Delta_h^*)^2\right]\mu^*
\left( w_r^*+u_z^*\right)
=\Delta_h^*\gamma_z^* ,
\end{equation}
where $\vartheta^*=\frac{1}{2\Delta_h^*}\left[
\frac{1}{a^*+h^*} -\frac{h^*_{zz}}{(\Delta_h^*)^2}
    \right]$ is the mean curvature and 
$\Delta_h^*=\sqrt{1+(h_z^*)^2}$. Considering the application of a radial electric field, a non-ionic insoluble surfactant is present at the gas-liquid interface of the film in region I, which reduces the surface tension of the liquid. Deformation of the film interface and variations in the interfacial velocity redistribute the insoluble surfactant along the liquid film surface, resulting in a non-uniform concentration field. The associated variation in surface tension generates Marangoni stresses and thereby alters the flow dynamics of the liquid film. The surfactant transport along the deformable interface is governed by the following convection-diffusion equation\cite{Gao2026}:
\begin{equation}
\label{eq:surfactant transport}
    \Gamma_t^*+w^*\Gamma_z^*
    +\frac{\Gamma^*}{(a^*+h^*)\Delta_h^*}
    \left((a^*+h^*)u_t^*\right)_z
    +\Gamma^*\vartheta^*u_n^*
    =
    \frac{D^*}{(a^*+h^*)\Delta_h^*}
    \left(
    \frac{(a^*+h^*)\Gamma_z^*}{\Delta_h^*}
    \right)_z ,
\end{equation}
 The scalar form used here is obtained by expanding the general vectorial transport equation in the coordinate system adopted in the present study, and the general formulation can be found in \citep{Stone1990,Wong1996,Kas-Danouche2002}. Here $D^*$ is the surface diffusivity, and $(u_n^*,u_t^*)$ are the normal and tangential components of interfacial velocity, which can be expressed as follows:
\begin{equation}
\label{eq:interfacial velocity}
    u_t^*=\frac{w^*+h_z^*u^*}{\Delta_h^*},
    \qquad
    u_n^*=\frac{u^*-h_z^*w^*}{\Delta_h^*}.
\end{equation}
To close the system of equations, assuming the fluid to be isothermal, a relation between the surface tension coefficient $\gamma^*$ and the interfacial surfactant concentration $\Gamma^*$ needs to be specified using the Langmuir equation of state \cite{Pereira2008}:
\begin{equation}
\label{eq:Langmuir}
\gamma^*=\gamma_0^*+R^*T^*\Gamma_\infty^*
    \ln\left(1-\frac{\Gamma^*}{\Gamma_\infty^*}\right),
\end{equation}
where $\gamma_0^*$, $R^*$ and $T^*$ correspond to surface tension of a clean liquid surface (without surfactant), universal gas constant and thermodynamic temperature. Equation (\ref{eq:laplace-potential}--\ref{eq:Langmuir}) form the dimensional governing system for the two-dimensional electrohydrodynamic flow of a surfactant-laden liquid film coating a vertical fiber.

The dimensional parameters are chosen with reference to the experiments of \citet{Kliakhandler2001}, which provide a classical data set for relatively thick films flowing down vertical fibers. Among the reported cases, the flow rate $Q^*=11.5~\mathrm{mm^3/s}$ is selected as the reference state. At this flow rate, the Rayleigh-Plateau bead train is regular and well developed, making it suitable for comparison with the traveling wave solutions discussed later. The electric field is formulated under the perfectly conducting liquid assumption used by \citet{Ding2014}, this assumption is relevant to practical coating liquids based on water mixtures, which usually have relatively high electrical conductivity. Their liquid-gas surface tension, however, is also relatively large. Experiments on films flowing down fibers have shown that large surface tension can promote the loss of axial symmetry of beads on fibers~\citep{Gabbard2021Asymmetric}. For this reason, the fluid properties used here are kept the same as those in the experiments of \citet{Kliakhandler2001}, while the electric field is treated in the same perfectly conducting limit as in \citet{Ding2014}. The surfactant-related parameters are chosen from commonly used ranges of maximum surface concentration and surface diffusivity, rather than from one specific surfactant formulation. For oil-based liquids, conventional hydrocarbon surfactants usually have only a limited ability to reduce the oil-air surface tension. Substantial reductions have nevertheless been reported for dioctyl sulfosuccinate sodium salt (AOT), sorbitan monooleate (Span 80), and surfactants containing fluorinated or silicone groups at oil-air or organic-liquid--air interfaces~\citep{Xu2017FoodOilSurfaceTension,Sawada2005FluoroalkylCooligomers}. Using one particular surfactant system would unnecessarily narrow the accessible parameter space. In the present formulation, the surfactant is therefore assumed to be insoluble, with adsorption and desorption neglected on the timescale of the film instability. This assumption isolates the Marangoni effect associated with interfacial transport. The dimensional physical parameters adopted in the present study are summarized in Table~\ref{tab1}.

\begin{table}
\centering
\setlength{\tabcolsep}{5pt}
\renewcommand{\arraystretch}{1.15}
\begin{tabular}{llll}
\toprule
\textbf{Parameter} & \textbf{Description} & \textbf{Value} & \textbf{Unit} \\
\midrule
$\mu^*$    & Dynamic viscosity        & 0.423              & $\mathrm{Pa\cdot s}$ \\
$\gamma^*$ & Surface tension          & $20$--$40$          & $\mathrm{dyn/cm}$ \\
$\rho^*$   & Density                  & 0.961              & $\mathrm{g/cm^3}$ \\
$g^*$      & Gravitational acceleration & 9.81             & $\mathrm{m/s^2}$ \\
$T^*$      & Thermodynamic temperature  & 298.15           & $\mathrm{K}$ \\
$R^*$      & Gas constant             & 8.31               & $\mathrm{N\cdot m/(K\cdot mol)}$ \\
$\Gamma_\infty^*$ & Maximum surfactant concentration & $10^{-10}$--$10^{-9}$ & $\mathrm{mol/cm^2}$ \\
$D^*$      & Surface diffusivity      & $10^{-6}$--$10^{-4}$ & $\mathrm{cm^2/s}$ \\
$a^*$      & Fiber radius             & 0.25               & $\mathrm{mm}$ \\
$Q^*$      & Volumetric flow rate     & 11.5               & $\mathrm{mm^3/s}$ \\
$\phi_0^*$ & Electric potential       & $0$--$3.03$         & $\mathrm{kV}$ \\
\bottomrule
\end{tabular}
\caption{Physical parameters used in the present study \citep{Kliakhandler2001,Li2023,Boulogne2012Instability}.}
\label{tab1}
\end{table}

\section {Scaling and reduced model.}
\label{Scaling and reduced model}

\subsection{Scaling and dimensionless formulation.}
\label{Scaling and dimensionless formulation.}
Following \citet{Kliakhandler2001}, \citet{Craster2006}, and \citet{Ding2014}, 
the radial length scale is taken to be the outer radius of the undisturbed film, $H=a^{*}+h_0^{*}$, where $a^{*}$ is the fiber radius and $h_0^{*}$ is the initial film thickness. Since the imposed control parameter in experiments is typically the volumetric flow rate $Q^{*}$, $h_0^{*}$ is obtained from the corresponding steady base flow. For a uniform film, whose interface is located at $r^{*}=a^{*}+h_0^{*}$, the base flow reduces to unidirectional Poiseuille-type flow driven by gravity. The axial velocity $w^{*}(r^{*})$ follows from the balance between gravitational forcing and radial viscous diffusion, with no slip at the fiber and zero shear at the free surface. The resulting axial velocity is
\begin{equation}
\label{eq: dimensional axial velocity}
w^{*}(r^{*})=
\frac{\rho^{*}g^{*}}{4\mu^{*}}
\left[
2\left(a^{*}+h_0^{*}\right)^2
\ln\left(\frac{r^{*}}{a^{*}}\right)
-\left((r^{*})^2-(a^{*})^2\right)
\right].
\end{equation}
The volumetric flow rate is then obtained by integrating this velocity profile over 
the annular film cross-section that $
Q^{*}=2\pi\int_{a^{*}}^{a^{*}+h_0^{*}} 
r^{*}w^{*}(r^{*})\,\mathrm{d}r^{*}.$
This yields the relation between the prescribed flow rate $Q^{*}$ and the initial film thickness $h_0^{*}$:
\begin{equation}
\label{eq: dimensional flow rate}
Q^*
=\frac{\pi \rho^* g^*}{8\mu^*}
\bigg\{
4\left(h_0^*+a^*\right)^4
\ln\left(\frac{h_0^*+a^*}{a^*}\right)  
-h_0^*
\left[
3\left(h_0^*\right)^3
+12a^*\left(h_0^*\right)^2
+14h_0^*\left(a^*\right)^2
+4\left(a^*\right)^3
\right]
\bigg\}.
\end{equation}
Following the work of Craster and Matar \cite{Craster2006}, we choose the length $\mathcal{L}=\gamma_0^*/(\rho^* g^* H)$ as the axial characteristic length scale. The remaining non-dimensional parameters are defined as follows:
\begin{equation}
\label{eq:non-dimensional parameters}
r=\frac{r^*}{H},\quad
z=\frac{z^*}{\mathcal{L}},\quad
\phi=\frac{\phi^*}{\phi_0^*},\quad
\gamma=\frac{\gamma^*}{\gamma_0^*},\quad
\Gamma=\frac{\Gamma^*}{\Gamma_\infty^*}, \quad
w=\frac{w^*}{W^*},\quad
u=\frac{u^*}{\epsilon W^*},\quad
t=\frac{t^*W^*}{\mathcal{L}},\quad
p=\frac{p^*-p_g^*}{\rho^*g^*\mathcal{L}} .
\end{equation}
where $\varepsilon=H/\mathcal{L},W^*=\rho^* g^* H^2/\mu^*$. Substituting the dimensionless variables defined in equation (\ref{eq:non-dimensional parameters}) into the original model, the electric potential equation is non-dimensionalized as follows:
\begin{equation}
\label{eq: nondimensional laplace-potential}
\phi_{rr}+r^{-1}\phi_r+\varepsilon^2\phi_{zz}=0,
\end{equation}
with $\phi=0$ at $r=\alpha+h(z,t)$ and $\phi=1$ at $r=\beta$.
Subsequently, the continuity and momentum equations are non-dimensionalized as follows:
\begin{equation}
u_r+r^{-1}u+w_z=0,
\label{eq:nondimensional continuity}
\end{equation}
\begin{equation}
\label{eq:nondimensional radial momentum}
\varepsilon^4 Re\left(u_t+u u_r+w u_z\right)
=-p_r+\varepsilon^2
\left(u_{rr}+r^{-1}u_r+u_{zz}-r^{-2}u\right),
\end{equation}
\begin{equation}
\label{eq:nondimensional axial momentum}
\varepsilon^2 Re\left(w_t+u w_r+w w_z\right)
=1-p_z+\left(w_{rr}+r^{-1}w_r+\varepsilon^2 w_{zz}\right).
\end{equation}
The boundary conditions are also written in non-dimensionalized form. At the fiber surface $r=\alpha$, the no-slip and no-penetration conditions give:
\begin{equation}
\label{nondimensional boundary:fiber}
u(\alpha)=w(\alpha)=0,
\end{equation}
At the free surface $r=\alpha+h(z,t)$, the kinematic condition and the normal and tangential stress balances are non-dimensionalized as follows, respectively:
\begin{equation}
\label{nondimensional boundary:kinematic}
u=h_t+w h_z,
\end{equation}
\begin{equation}
\label{nondimensional boundary:normal stress}
\Delta_h^2 p-2\varepsilon^2\left[u_r-h_z\left(w_r+\varepsilon^2 u_z\right)
+\varepsilon^2 h_z^2 w_z
\right]  +E_b\left[\frac{1}{2}
\left(1-\varepsilon^2 h_z^2\right)
\left(\phi_r^2-\varepsilon^2\phi_z^2\right)
-2\varepsilon^2 h_z\phi_r\phi_z
\right]
=2\varepsilon Bo^{-1}\Delta_h^2\gamma\vartheta ,
\end{equation}
\begin{equation}
\label{nondimensional boundary:tangential stress}
2\varepsilon^2 h_z\left(u_r-w_z\right)
+\left(2-\Delta_h^2\right)
\left(\varepsilon^2 u_z+w_r\right)
=\varepsilon Bo^{-1}\Delta_h\gamma_z .
\end{equation}

The convection-diffusion equation for the surfactant concentration is non-dimensionalized in the following manner:
\begin{equation}
\label{nondimensional eq:surfactant transport}
\Gamma_t+w\Gamma_z
+\frac{\Gamma}{(\alpha+h)\Delta_h}
\left[\frac{(\alpha+h)(w+\varepsilon^2 u h_z)}{\Delta_h}
\right]_z
+\Gamma\vartheta\frac{u-w h_z}{\Delta_h}  
=\frac{1}{(\alpha+h)\Delta_h Pe}
\left[\frac{(\alpha+h)\Gamma_z}{\Delta_h}
\right]_z ,
\end{equation}
and
where $\vartheta=\frac{1}{2\Delta_h}\left(\frac{1}{\alpha+h}
-\frac{\varepsilon^2 h_{zz}}{\Delta_h^2}\right)$ is the non-dimensional mean curvature, and $\Delta_h=\sqrt{1+\varepsilon^2 h_z^2}$.
The relationship between the non-dimensional surface tension coefficient and the concentration of surfactant is
\begin{equation}
\label{nondimensional eq:Langmuir}
\gamma=1+Ma\ln(1-\Gamma).
\end{equation}
The non-dimensional parameters derived from the governing equations and boundary conditions are presented as follows:
\begin{equation}
\alpha=\frac{a^*}{H},\quad
\beta=\frac{b^*}{H},\quad
Pe=\frac{W^*\mathcal{L}}{D^*},\quad
Re=\frac{\rho^*W^*\mathcal{L}}{\mu^*},\quad
Bo=\frac{\rho^*g^*H^2}{\gamma_0^*},\quad
Ma=\frac{R^*T^*\Gamma_\infty^*}{\gamma_0^*},\quad
E_b=\frac{\epsilon\varepsilon(\phi_0^*)^2}{\rho^*g^*H^3}.
\end{equation}
The non-dimensional fiber radius is denoted by $\alpha$, a smaller value of $\alpha$ corresponds to a thicker liquid film. The non-dimensional radius of the outer electrode is denoted by $\beta$. The remaining non-dimensional groups are the Péclet number $Pe$, Reynolds number $Re$, Bond number $Bo$, Marangoni number $Ma$, and electric Weber number $E_b$. In the next section, a reduced model is derived using the long wave approach of Ding \textit{et al.}~\cite{Ding2014} and Craster and Matar~\cite{Craster2006}. The small parameter is $\varepsilon=H/\mathcal{L}$, which measures the ratio of the radial length scale of the liquid film to the axial capillary length scale. The Bond number measures the relative importance of gravity and capillarity. With the present axial length scale $\mathcal{L}=\gamma_0^*/(\rho^*g^*H)$, one obtains $Bo=H/\mathcal{L}=\varepsilon$. Hence, in the present scaling, the Bond number is fixed by the long wave parameter and is not treated as an independent control parameter. The Péclet number measures the relative strength of convection along the interface and surface diffusion in the surfactant transport. Estimates based on Table~\ref{tab1} give $Pe\gg1$, indicating that interfacial convection is much stronger than surface diffusion, and a representative value of $Pe=10^3$ is used in the following calculations. The Reynolds number measures the relative strength of inertia and viscous effects in the liquid film. For the present parameter set, $Re=[0.09,0.19]$, which is of the same order as $\varepsilon$. The Marangoni number measures the magnitude of surface tension variations caused by surfactant transport relative to the surface tension of a clean interface. From the ranges of $\Gamma_\infty^*$ and $\gamma_0^*$ in Table~\ref{tab1}, the estimated physical range is $Ma=[0.06,1.24]$, showing that $Ma$ is of order one. Given that the case without surfactants corresponds to $Ma=0$, the parameter range of $Ma$ is set as $Ma=[0,1]$ in the present study. The electric Weber number measures the strength of the electric forcing relative to gravity. The applied electric potential listed in Table~\ref{tab1} corresponds to $E_b=[0,3]$, showing that $E_b$ is also of order one. Unless otherwise stated, $\alpha=0.2856$ and $\varepsilon=0.233$ which correspond to the flow rate associated with regime $b$ in the experiments of \citet{Kliakhandler2001}, are adopted throughout the following analysis. The main control parameters examined below are $Ma$ and $E_b$, which describe the effects of Marangoni stresses and electric stresses on the instability of the coating film.

\subsection{Derivation of the one-dimensional model.}
\label{Derivation of the one-dimensional model.}
By extracting the leading-order term of $\varepsilon$, the electric potential equation is simplified as follows:
\begin{equation}
\label{reduced electric potential}
\phi_{rr}+r^{-1}\phi_r=0,
\end{equation}
with $\phi=0$ at $r=\alpha+h(z,t)$ and $\phi=1$ at $r=\beta$. Thus, the solution of the electric potential equation can be obtained as follows:
\begin{equation}
\label{reduced potential solution}
\phi=\frac{\ln(\alpha+h)-\ln r}{\ln(\alpha+h)-\ln\beta}.
\end{equation}

Moreover, the radial momentum equation and normal boundary condition are reduced to:
\begin{equation}
\label{reduced radial momentum}
p_r=0,
\end{equation}
\begin{equation}
\label{reduced normal boundary}
p+\frac{E_b}{2}\phi_r^2
=\gamma\left(
\frac{1}{\alpha+h}
-\varepsilon^2 h_{zz}
\right),
\end{equation}
which can derive the expression for the pressure as follows:
\begin{equation}
\label{reduced pressure solution}
p=-\frac{E_b}
{2\left[\ln(\alpha+h)-\ln\beta\right]^2(\alpha+h)^2}
+\gamma\left(\frac{1}{\alpha+h}
-\varepsilon^2h_{zz}
\right),
\end{equation}
where $\gamma=1+Ma\ln(1-\Gamma)$. The term $\varepsilon^2 h_{zz}$ in equation (\ref{reduced pressure solution}) is retained because it accounts for the axial curvature contribution to the capillary pressure~\cite{Craster2006}. Although formally of higher order relative to the leading circumferential curvature contribution, this term is essential for capturing the stabilization of disturbances with relatively short wavelengths. Without this term, the model cannot correctly capture the cutoff wave number of the Rayleigh-Plateau instability. The simplified axial momentum equation, along with the no-slip boundary condition and the tangential stress balance condition, are as follows:
\begin{equation}
\label{reduced axial momentum}
1-p_z+w_{rr}+r^{-1}w_r=0,
\end{equation}
\begin{equation}
\label{reduced tangential stress}
w(\alpha)=0,
\qquad
w_r\big|_{r=\alpha+h(z,t)}=\gamma_z .
\end{equation}

Therefore, by applying boundary condition above to solve the axial momentum equation, the expression for axial velocity is obtained:
\begin{equation}
\label{reduced axial velocity}
w=\frac{r^2-\alpha^2}{4}(p_z-1)+
\left[\frac{(\alpha+h)^2}{2}(1-p_z)
+(\alpha+h)\gamma_z\right]
\ln\frac{r}{\alpha}.
\end{equation}

The leading-order equations for the continuity equation and kinematic boundary conditions are as follows:
\begin{equation}
\label{reduced continuity}
u_r+r^{-1}u+w_z=0,
\end{equation}
\begin{equation}
\label{reduced kinematic}
u=h_t+wh_z.
\end{equation}

Combining the continuity equation, the integral form of the kinematic boundary condition is rewritten as follows:
\begin{equation}
\label{reduced flow-rate}
h_t+(\alpha+h)^{-1}q_z=0,
\end{equation}
where $q$ is the flow rate. For notational convenience, the flow rate is written as
\begin{equation}
q=-\frac{p_z-1}{4}\mathcal{F}(\alpha,h)+\mathcal{G}(\alpha,h)\gamma_z,
\end{equation}
with
\begin{equation}
\mathcal{F}(\alpha,h)=(\alpha+h)^4\ln\left(\frac{\alpha+h}{\alpha}\right)
-\frac{h(2\alpha+h)(2\alpha^2+6\alpha h+3h^2)}{4},
\end{equation}
and
\begin{equation}
\mathcal{G}(\alpha,h)=\frac{(\alpha+h)^3}{2}
\ln\left(\frac{\alpha+h}{\alpha}\right)
-\frac{h(\alpha+h)(2\alpha+h)}{4}.
\end{equation}

Then the convection-diffusion equation for surfactant concentration can be simplified to the following form:
\begin{equation}
\label{reduced convection-diffusion}
\Gamma_t+(w\Gamma)_z+\frac{\Gamma u}{\alpha+h}
=\frac{1}{(\alpha+h)Pe}
\left[(\alpha+h)\Gamma_z\right]_z .
\end{equation}
Thus, Eqs.~(\ref{reduced flow-rate})--(\ref{reduced convection-diffusion}) therefore define the reduced governing system. The pressure is given by Eq.~(\ref{reduced pressure solution}), and the axial velocity $w$ at the interface is obtained from Eq.~(\ref{reduced axial velocity}) by setting $r=\alpha+h$. The radial velocity $u$ at the interface can be obtained from the kinematic boundary condition, Eq.~(\ref{reduced kinematic}), or equivalently from the continuity equation Eq.~(\ref{reduced continuity}). In the following discussion, unless otherwise stated, $u$ and $w$ denote the velocities evaluated at the interface.

\section{Linear stability and mode competition.}
\label{Linear stability and mode competition}
At the early stage of the flow evolution, the disturbance amplitude remains small compared with the basic flow, allowing the initial evolution of the falling film to be examined using linear stability analysis. The quantities in the basic state are given by:
\begin{equation}
\label{DR: base flow}
\left\{
\begin{aligned}
  & \bar{\gamma} = 1 + Ma \ln \left( 1 - \bar{\Gamma} \right),\quad \bar{u} = 0,\quad \bar{h} =1-\alpha, \\
  & \bar{p} = \bar{\gamma}-\frac{E_b}{2(\ln \beta)^2}, \quad \bar{\mathcal{G}}=\bar{w} = \dfrac{1}{4} \left( \alpha^2 - 1 - 2 \ln \alpha \right), \\
  & \bar{\mathcal{F}} = - \ln \alpha - \dfrac{1}{4} \left( 1 - \alpha^2 \right) \left( 3 - \alpha^2 \right), \bar{\Gamma}=0.4.
\end{aligned}
\right.
\end{equation}
Following Gao \textit{et al.} \cite{Gao2026} and Li \textit{et al.} \cite{Li2023}, the initial interfacial concentration of the surfactant $\bar{\Gamma}$ is set to $0.4$ times the maximum interfacial concentration, which allows the Marangoni effect to be more clearly reflected.
It should be noted that surfactants at sufficiently high concentrations may substantially enhance interfacial viscosity\citet{Yang_2025}. The present study focuses primarily on the coupling between Marangoni stresses and the electric field, and the effects of interfacial viscosity are therefore not included in the current model.
 The physical quantities are then expressed as the sum of a basic state and a small disturbance:
\begin{equation}
\label{DR: perturbations assumption}
[u,w,p,h,\Gamma,\gamma,\mathcal{F},\mathcal{G}]=[\bar{u},\bar{w},\bar{p},\bar{h},\bar{\Gamma},\bar{\gamma},\bar{\mathcal{F}},\bar{\mathcal{G}}]+[\hat{u},\hat{w},\hat{p},\hat{h},\hat{\Gamma},\hat{\gamma},\hat{\mathcal{F}},\hat{\mathcal{G}}]e^{ik(z-ct)}+c.c..
\end{equation}
Here, $k$ is the wave number and $c=c_r+i c_i$ is the complex wave speed, whose imaginary part $c_i$ determines the stability of the disturbance: $c_i>0$ corresponds to temporal growth and hence an unstable state, whereas $c_i\leq0$ indicates that the disturbance does not grow and the system remains stable. The real part $c_r$ gives the propagation speed of the disturbance, and $c.c.$ denotes the complex conjugate. The corresponding disturbance amplitudes are then given by:
\begin{equation}
\label{DR: perturbations solution}
\left\{
\begin{aligned}
  \hat{\gamma}&=\frac{Ma}{\bar{\Gamma} - 1} \hat{\Gamma},
  \quad  \hat{\mathcal{F}}= 8\bar{w}\hat{h}, \quad \hat{\mathcal{G}} = (\bar{w}-\ln \alpha)\hat{h} \\
  \hat{p} & =\frac{E_b(\ln \beta-1)}{(\ln \beta)^3}\hat{h} -\bar{\gamma} \left( 1 - \varepsilon^2 k^2 \right) \hat{h}
  + \hat{\gamma}, \\
  \hat{w} & = - i k \bar{w} \hat{p} - \ln \alpha( \hat{h}
  + i k\hat{\gamma}).
\end{aligned}
\right.
\end{equation}

The relations above allow all other disturbance quantities to be expressed in terms of $\hat{h}$ and $\hat{\Gamma}$, which can therefore be taken as the independent disturbance amplitudes. Substituting these relations into the linearized equation governing the film thickness and the transport equation for the surfactant gives the following two algebraic equations:
\begin{equation}
\label{DR: film thickness}
-ikc\hat{h}+2ik\bar{w}\hat{h}+\frac{1}{4}{{k}^{2}}\bar{\mathcal{F}}\hat{p}-k^2\bar{w}\hat{\gamma}=0,
\end{equation}
\begin{equation}
\label{DR: surfactant transport}
-ikc\hat{\Gamma }+ik\bar{w}\hat{\Gamma }+ik\bar{\Gamma }\hat{w}+\bar{\Gamma }\left( -ikc\hat{h}+ik\bar{w}\hat{h} \right)+\frac{1}{Pe}{{k}^{2}}\hat{\Gamma }=0.
\end{equation}
The disturbance amplitude $\hat{u}$ at the free surface has been eliminated using the linearized kinematic boundary condition, leaving a closed system for $\hat{h}$ and $\hat{\Gamma}$. Requiring this system to admit a nontrivial solution then yields the dispersion relation between the wave number $k$ and the complex wave speed $c$. Since the resulting relation is algebraic in $c$, an analytical solution can be derived, although its full expression is too cumbersome to reveal clearly how the physical parameters affect the flow stability, which motivates the following expansion in the long wave limit.
\subsection{Asymptotic solution in the long wavelength limit }
\label{Asymptotic solution in the long wavelength limit}
Following the asymptotic procedure adopted by Gao \textit{et al.}~\cite{Gao2026}, the wave number is assumed to satisfy $k\ll1$, allowing the disturbance amplitudes and the complex wave speed to be expanded as
\begin{equation}
\label{As: expansion}
\left(\hat{h},\hat{\Gamma},c\right)
=\left(\hat{h}_0,\hat{\Gamma}_0,c_0\right)
+k\left(\hat{h}_1,\hat{\Gamma}_1,c_1\right)
+O\left(k^2\right).
\end{equation}
Since the stability of the system is determined by $c_i$, the coefficients in the expansion of the complex wave speed $c$ are determined successively in the long wavelength limit, with the equations at the leading order given by:
\begin{equation}
\label{As: leading order h}
(2\bar{w}-c_0)\hat{h}_0=0,
\end{equation}
\begin{equation}
\label{As: leading order Gamma}
(\bar{w}-c_0)\hat{\Gamma}_0
+\bar{\Gamma}\left(\bar{w}-c_0-\ln\alpha\right)\hat{h}_0
=0.
\end{equation}
Equations~\eqref{As: leading order h} and~\eqref{As: leading order Gamma} reveal two distinct disturbance modes. For the mode associated with disturbance in surfactant concentration, $\hat{h}_0=0$ and $\hat{\Gamma}_0\neq0$, giving $c_0=\bar{w}$, which indicates that the disturbance propagates downstream at the interfacial velocity of the basic state. When the interface disturbance is nonzero, $\hat{h}_0\neq0$, the solution becomes $\hat{\Gamma}_0=-[(\ln\alpha+\bar{w})\bar{\Gamma}/\bar{w}]\hat{h}_0,$ and $c_0=2\bar{w},$ showing that this mode propagates downstream at twice the interfacial velocity of the basic state. These two branches correspond to the Marangoni mode associated with variations in surfactant concentration and the Rayleigh--Plateau mode associated with interface disturbance, in agreement with the results of Gao \textit{et al.}~\cite{Gao2026}. Since the solutions at the leading order determine only the propagation speeds of the two modes, their stability must be established from the imaginary part of the complex wave speed obtained at the subsequent orders. The terms at the first order $O(k)$ in Eqs.~\eqref{DR: film thickness} and~\eqref{DR: surfactant transport} are therefore considered below:
\begin{equation}
\label{As: first order h}
(c_0-2\bar{w})\hat{h}_1+\hat{h}_0c_1
-i\frac{Ma\bar{w}}{\bar{\Gamma}-1}\hat{\Gamma}_0 
+\frac{i}{4}\bar{\mathcal{F}}
\left[
\frac{E_b(\ln\beta-1)}{(\ln\beta)^3}\hat{h}_0
-\left(1-\varepsilon^2k^2\right)\bar{\gamma}\hat{h}_0
+\frac{Ma}{\bar{\Gamma}-1}\hat{\Gamma}_0
\right]
=0 .
\end{equation}
\begin{equation}
\label{As: first order Gamma}
\begin{aligned}
&\left(\hat{\Gamma}_0+\bar{\Gamma}\hat{h}_0\right)c_1
+(c_0-\bar{w})\hat{\Gamma}_1 
+\left(\ln\alpha-\bar{w}+c_0\right)\bar{\Gamma}\hat{h}_1 \\
&\quad
+i\bar{w}\bar{\Gamma}\left[
\frac{E_b(\ln\beta-1)}{(\ln\beta)^3}\hat{h}_0
-\left(1-\varepsilon^2k^2\right)\bar{\gamma}\hat{h}_0
+\frac{Ma}{\bar{\Gamma}-1}\hat{\Gamma}_0
\right] +i\ln\alpha\,
\frac{Ma\bar{\Gamma}}{\bar{\Gamma}-1}\hat{\Gamma}_0
+\frac{i}{Pe}\hat{\Gamma}_0=0 .
\end{aligned}
\end{equation}
It is worth noting that, unlike the treatment adopted by \citet{Gao2026}, the axial curvature term is retained when the equations at first order are derived, which is essential for distinguishing the two modes in the subsequent numerical results. As in the leading order system, the equations at first order also admit two solution branches. Substituting the leading order solution for the Rayleigh-Plateau mode associated with interface disturbance directly into Eq.~\eqref{As: first order h} gives the following expression for $c_1$:
\begin{equation}  
\label{As: RP solution}
{{c}_{1}}=
\underbrace{i\frac{1}{4}\bar{F}\frac{E_b(1-\ln \beta)}{(\ln\beta)^3}}_{\text{Effect of electric stress}}
\quad+\quad
\underbrace{i\frac{1}{4}\bar{F} (1-\varepsilon^2k^2)
\bar{\gamma }}_{\text{Effect of capillarity}}
\quad-\quad
\underbrace{i\left( \frac{1}{4}\bar{F}-\bar{w} \right)\frac{Ma\bar{\Gamma}\left( \bar{w}+\ln \alpha  \right)}{\left( 1- \bar{\Gamma } \right)\bar{w}}}_{\text{Effect of Marangoni stress}}.
\end{equation}
The results show that, in the long wavelength limit, the stability of the Rayleigh-Plateau mode is governed jointly by electric stress, capillarity, and Marangoni stress. When $E_b=0$, the electric field is absent and the present result reduces to that obtained by Gao \textit{et al.}~\cite{Gao2026}. The influence of electric stress depends on the position of the outer electrode, represented by $\beta$, with electric stress enhancing the growth of disturbances for $\beta<e$ but suppressing it for $\beta>e$, in agreement with the results of Ding \textit{et al.}~\cite{Ding2014}. In contrast to electric stress, whose influence changes with $\beta$, capillarity always promotes instability, whereas Marangoni stress has a stabilizing effect. The stabilizing influence of $Ma$ arises through two mechanisms: increasing $Ma$ reduces the surface tension of the basic state, $\bar{\gamma}$, thereby weakening the instability caused by capillarity, while its direct contribution to Marangoni stress further suppresses the growth of disturbances. It is worth noting that, at the present order, the contribution from electric stress appears as an independent additive term and does not directly modify the contributions from capillarity or Marangoni stress.

For the second solution branch, substituting the leading order solution of the Marangoni mode associated with the disturbance in surfactant concentration into Eq.~\eqref{As: first order h} yields the first order solution for the interface disturbance:
\begin{equation}
\label{As: h1 solution}
\hat{h_1}=i(\frac{\mathcal{F}}{4\bar{w}}-1)\frac{Ma\hat{\Gamma}_0}{\bar{\Gamma}-1}.
\end{equation}
Further substitution of the above solution for the interface disturbance at first order into Eq.~\eqref{As: first order Gamma} gives the following expression for $c_1$ associated with the Marangoni mode:
\begin{equation}
\label{As: Maragoni solution}
c_1 =-\quad \underbrace{ i\left[ \bar{w} +  \frac{\bar{\mathcal{F}} \ln \alpha}{4 \bar{w}} \right] \frac{Ma \bar{\Gamma}}{\bar{\Gamma}-1} }_{\text{Effect of Marangoni stress}}\qquad  - \underbrace{ i \frac{1}{Pe} }_{\text{Effect of surface diffusion}}.
\end{equation}
Equation~\eqref{As: Maragoni solution} shows that the Marangoni mode is stable in the long wavelength limit, with its decay rate governed by both Marangoni stress and surface diffusion. Since $Pe$ measures the relative strength of convection and diffusion along the interface, a smaller value of $Pe$ corresponds to stronger surface diffusion, which increases the magnitude of the negative imaginary part of $c_1$ and therefore enhances the stabilization of the disturbance.
\subsection{Full solution for arbitrary wavenumbers}
\label{Full solution for arbitrary wavenumbers}
Although the asymptotic analysis above reveals the propagation characteristics and stability mechanisms of the two modes in the long wavelength limit, its validity is restricted to $k\ll1$ and cannot describe the stability of the system over the entire range of wave numbers, which is therefore examined by solving the full eigenvalue problem in terms of the complex temporal eigenvalue $\lambda=\lambda_r+i\lambda_i$, where $\lambda_r=kc_i$ is the temporal growth rate and $\lambda_i=-kc_r$ is the intrinsic frequency. Accordingly, $\lambda_r>0$ indicates instability, whereas $\lambda_r\leq0$ corresponds to a stable state. To obtain the temporal eigenvalue at arbitrary wave numbers, Eqs.\eqref{DR: film thickness} and \eqref{DR: surfactant transport} are solved simultaneously, yielding:
\begin{equation}
\label{fs: lambda}
\lambda=
\frac{-\left(\mathcal{A}_1+\mathcal{A}_3-\mathcal{A}_4\right)
\pm
\sqrt{\left(\mathcal{A}_1+\mathcal{A}_3-\mathcal{A}_4\right)^2
-4\left(\mathcal{A}_1\mathcal{A}_3-\mathcal{A}_2\mathcal{A}_4\right)}
}{2},
\end{equation}
where
\begin{equation}
\label{fs: A}
\left\{
\begin{aligned}
\mathcal{A}_1={}&ik\bar{w}
+k^2(\bar{w}+\ln\alpha)
\frac{Ma\bar{\Gamma}}{\bar{\Gamma}-1}
+\frac{k^2}{Pe},
\\[4pt]
\mathcal{A}_2={}&ik\bar{w}-ik\ln\alpha
+k^2\bar{w}
\left[
\frac{E_b(\ln\beta-1)}{(\ln\beta)^3}
-\bar{\gamma}\left(1-\varepsilon^2k^2\right)
\right],
\\[4pt]
\mathcal{A}_3={}&2ik\bar{w}
+\frac{k^2\bar{\mathcal{F}}}{4}
\left[
\frac{E_b(\ln\beta-1)}{(\ln\beta)^3}
-\bar{\gamma}\left(1-\varepsilon^2k^2\right)
\right],
\\[4pt]
\mathcal{A}_4={}&k^2\bar{\mathcal{F}}
\frac{Ma\bar{\Gamma}}{4(\bar{\Gamma}-1)}.
\end{aligned}
\right.
\end{equation}
When $E_b=0$, the present system reduces to the result obtained by \citet{Gao2026}. In the separate limit $Ma=0$, the dispersion relation formally yields two eigenvalues, $\lambda=-\mathcal{A}_1$ and $\lambda=-\mathcal{A}_3$. The former arises from the surfactant transport equation and becomes decoupled from the interfacial dynamics when the Marangoni effect vanishes, it is therefore excluded from the physically relevant dispersion relation. Consequently, the only physical solution is $\lambda=-\mathcal{A}_3$, which is consistent with the result of \citet{Ding2014}, in this case, whether the electric field stabilizes or destabilizes the system depends on the position of the outer electrode: the electric field is destabilizing when $\beta<e$, whereas it is stabilizing when $\beta>e$. If $E_b$ is further set to zero, the system reduces to the surfactant-free and electrically passive case considered by \citet{Craster2006}, for which the most unstable and cutoff $k_m=1/(\sqrt{2}\varepsilon), k_c=1/\varepsilon$,
respectively. It should be noted that the wavenumber and temporal growth rate introduced above are nondimensionalized using the capillary length. To facilitate comparison with previous studies, the rescaled quantities $\omega=\varepsilon\lambda, \kappa=\varepsilon k$, are adopted throughout the subsequent analysis.
\begin{figure}
    \centering
    \begin{subfigure}[b]{0.45\textwidth}
        \centering
        \includegraphics[width=\textwidth, trim={0pt 0pt 0pt 0pt}, clip]{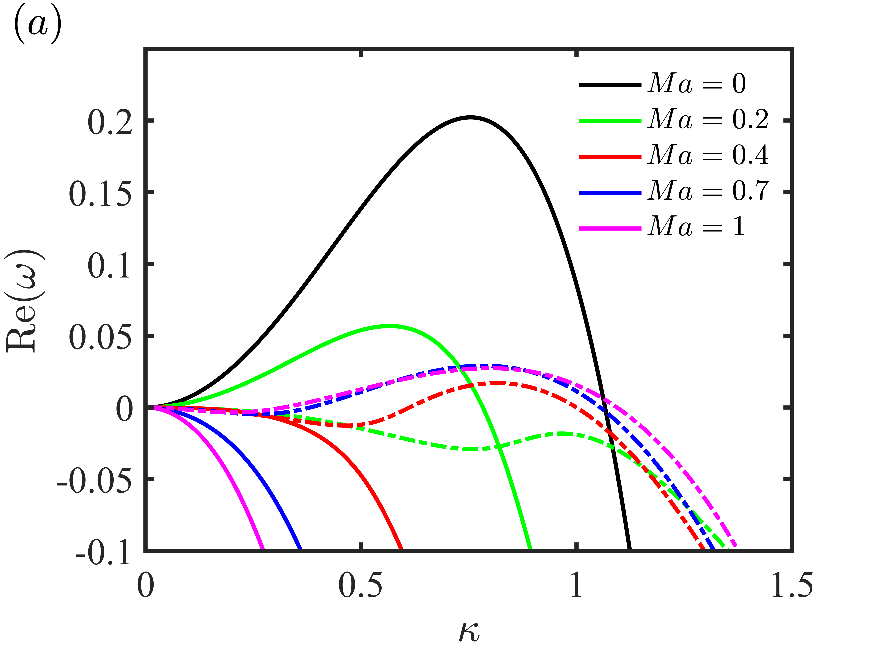}
    \end{subfigure}
    \hspace{0.02\textwidth}
    \begin{subfigure}[b]{0.45\textwidth}
        \centering
        \includegraphics[width=\textwidth, trim={0pt 0pt 0pt 0pt}, clip]{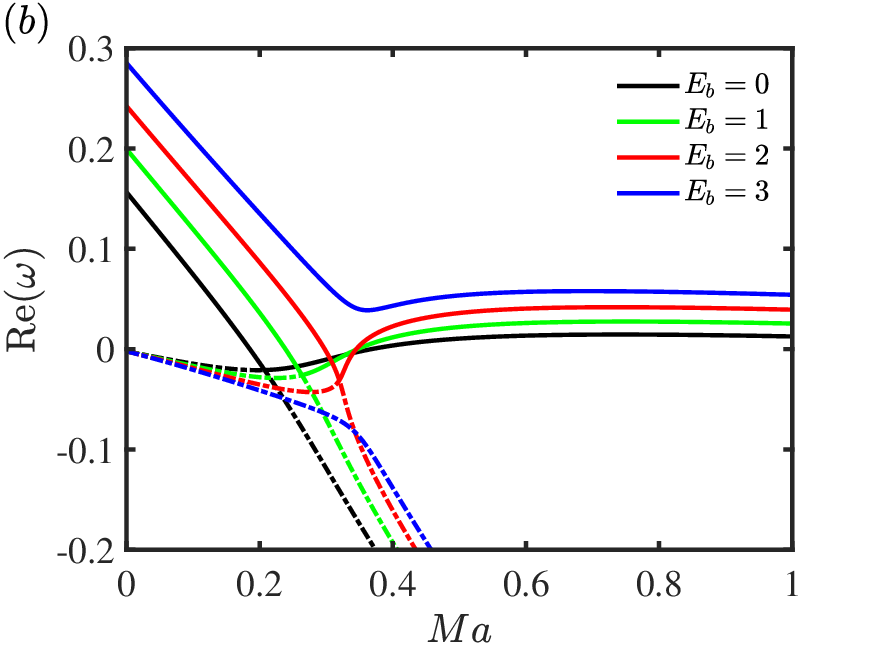}
    \end{subfigure}
       
    \caption{\raggedright
$(a)$ The temporal growth rate $\omega$ versus the wavenumber $\kappa$ under the combined effects of the Marangoni number $Ma$ and a radial electric field, with $E_b=1$ and $\beta=e^{0.9}$. The solid and dashed lines correspond to the RP and Marangoni modes, respectively. 
$(b)$ The temporal growth rate $\omega$ versus $Ma$ under a radial electric field, with $\beta=e^{0.9}$ and $\kappa=1/\sqrt{2}$. The solid and dash-dotted lines denote the larger and smaller values of $\omega$, respectively.}
    \label{fig2}
\end{figure}

To further elucidate the influence of the Marangoni number $Ma$ on the stability of a thick liquid film subjected to a radial electric field, the temporal growth rates of the system are examined in \autoref{fig2}. \autoref{fig2}($a$) presents the dispersion curves at different values of $Ma$ for $E_b=1$ and $\beta=e^{0.9}<e$. In the absence of surfactant, namely when $Ma=0$, only the classical Rayleigh-Plateau (RP) mode exists in the system. Once an insoluble surfactant is introduced, an additional Marangoni mode associated with interfacial transport emerges. The solid and dash-dotted lines represent the RP and Marangoni modes, respectively, in agreement with the results of \citet{Gao2026}, while the procedure used to distinguish the two modes is provided in Appendix \ref{appA}. As $Ma$ increases, both the maximum growth rate and the cutoff wavenumber of the RP mode decrease rapidly, demonstrating that the Marangoni effect effectively suppresses the RP instability. At relatively small values of $Ma$, the Marangoni mode remains stable. With a further increase in $Ma$, however, this mode becomes unstable and eventually replaces the RP mode as the dominant unstable mode. The maximum growth rate of the Marangoni mode first increases and then decreases with $Ma$, indicating that the instability induced by surfactant transport responds non-monotonically to the Marangoni effect. Nevertheless, its maximum growth rate remains lower than that of the surfactant-free RP mode, suggesting that the instability associated with the Marangoni mode is comparatively weak. Because both the electric field and the Marangoni effect alter the maximum growth rate and the most unstable wavenumber, the most unstable wavenumber of the reference state without either an electric field or surfactant, $\kappa=1/\sqrt{2}$ is adopted in the following analysis so that the influence of each mechanism can be examined separately. \autoref{fig2}($b$) shows the variation of the two temporal eigenvalue branches with $Ma$ at $\beta=e^{0.9}$. The solid and dash-dotted lines denote the larger and smaller growth rates, respectively, while $\mathrm{Re}(\omega)$ represents the real part of the corresponding eigenvalue. When $E_b<3$, the larger growth rate initially decreases as $Ma$ increases, becomes negative over an intermediate range of $Ma$, and subsequently rises above zero again. By contrast, the smaller growth rate remains negative throughout the parameter range considered. The non-monotonic variation of the upper branch does not represent the simple evolution of a single mode, but instead results from competition and an exchange of dominance between the two modes. At small $Ma$, the RP mode is dominant and is progressively suppressed by the Marangoni effect, whereas beyond the turning point, the Marangoni mode gradually becomes the dominant source of instability. Consequently, a finite interval of $Ma$ exists within which both growth rates are negative, so that the disturbance at the selected wavenumber is completely suppressed. Because the radial electric field is destabilizing when $\beta<e$, increasing $E_b$ raises the value of $Ma$ required to enter the stable regime and progressively narrows the stability window. Notably, the growth rate at the transition from the RP-dominated branch to the Marangoni-dominated branch decreases continuously as $E_b$ increases. When $E_b=3$, the turning point disappears and both branches become smooth, rather than exhibiting the distinct junction between the RP-dominated and Marangoni-dominated portions observed at lower $E_b$.
\begin{figure}
    \centering
    \begin{subfigure}[b]{0.45\textwidth}
        \centering
        \includegraphics[width=\textwidth,
      trim={0pt 0pt 0pt 0pt}, clip]{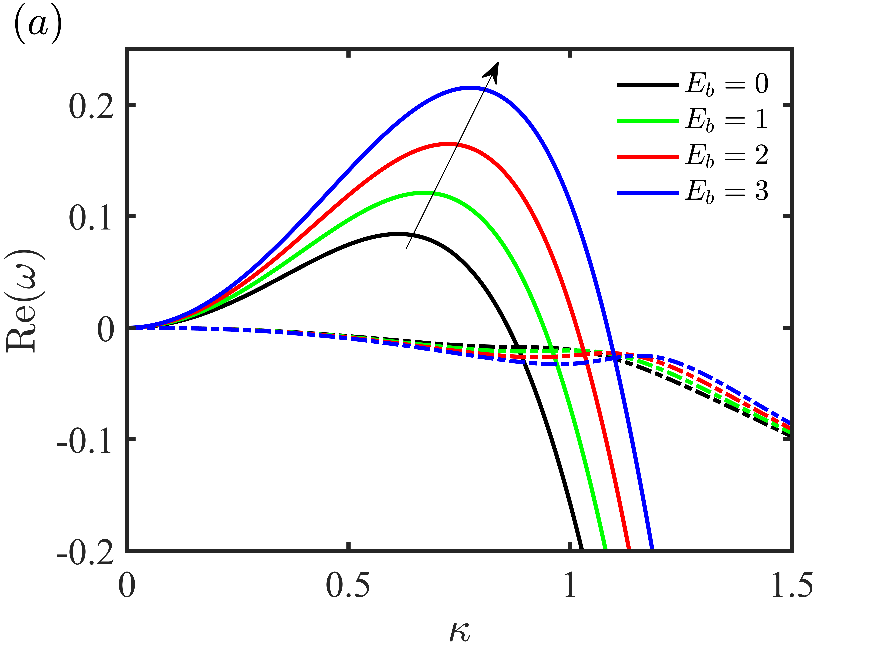}
    \end{subfigure}
    \hspace{0.02\textwidth}
    \begin{subfigure}[b]{0.45\textwidth}
        \centering
        \includegraphics[width=\textwidth,
        trim={0pt 0pt 0pt 0pt},clip]{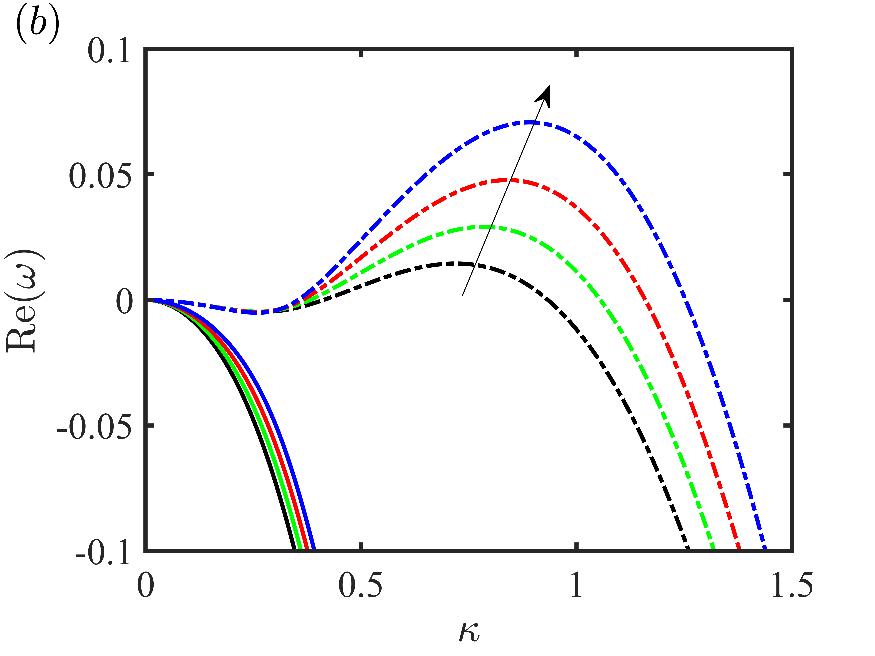}
    \end{subfigure}
    \vspace{0.02\textwidth}
    \begin{subfigure}[b]{0.45\textwidth}
        \centering
        \includegraphics[width=\textwidth,
            trim={0pt 0pt 0pt 0pt},clip]{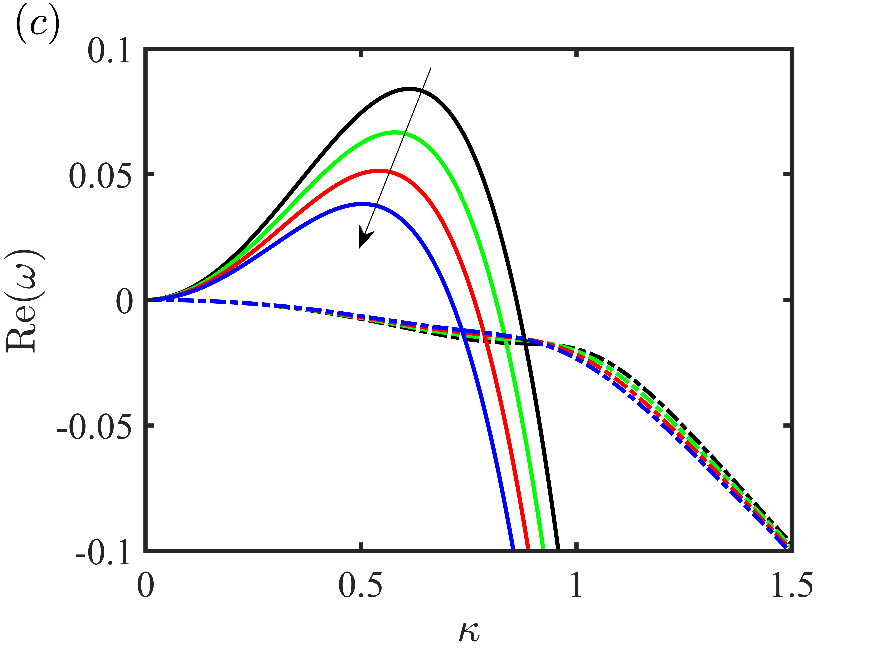}
    \end{subfigure}
    \hspace{0.02\textwidth}
    \begin{subfigure}[b]{0.45\textwidth}
        \centering
        \includegraphics[width=\textwidth,trim={0pt 0pt 0pt 0pt},clip]{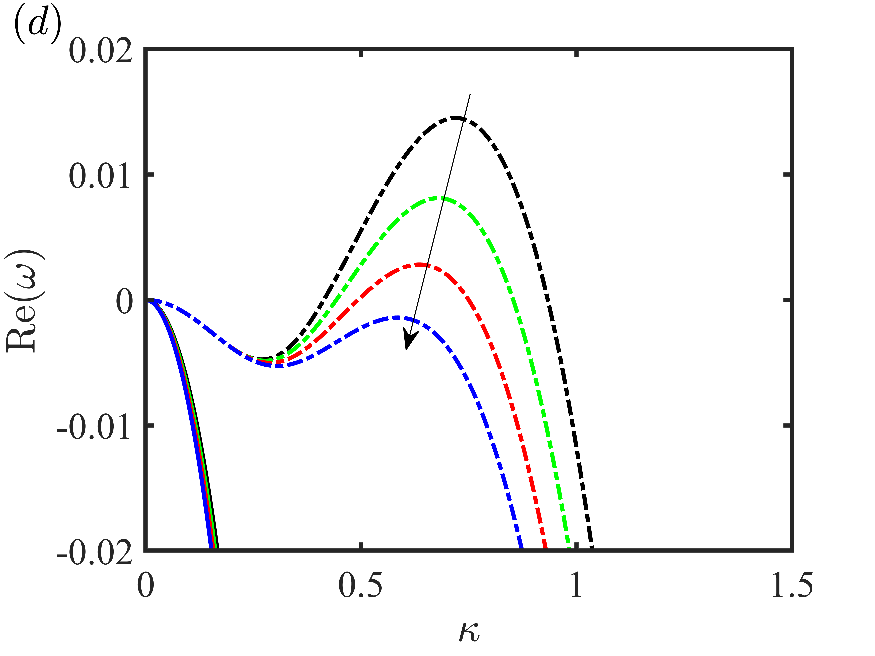}
    \end{subfigure}
    \caption{\raggedright The real temporal growth rate ($\omega$) versus the wavenumber $\kappa$ under the effects of $E_b$.
$(a)$ $Ma=0.1$, $\beta=e^{0.9}$.
$(b)$ $Ma=0.7$, $\beta=e^{0.9}$.
$(c)$ $Ma=0.1$, $\beta=e^{1.1}$.
$(d)$ $Ma=0.7$, $\beta=e^{1.1}$.The solid and dash-dotted lines represent the RP mode and the Marangoni mode, respectively.}
    \label{fig3}
\end{figure}

To further illustrate the influence of the electric field on the stability of the system, \autoref{fig3} presents the dependence of the temporal growth rate on the wavenumber at different values of $E_b$ for both $\beta<e$ and $\beta>e$. The case $Ma=0.1$ corresponds to a regime dominated by the RP mode, whereas $Ma=0.7$ corresponds to a regime dominated by the Marangoni mode. \autoref{fig3}($a$-$b$) shows the results for $\beta<e$. Regardless of whether the RP mode or the Marangoni mode is dominant, the maximum growth rate, the most unstable wavenumber, and the cutoff wavenumber all increase with $E_b$, demonstrating the destabilizing effect of the electric field. By contrast, for $\beta>e$, as shown in \autoref{fig3}($c$-$d$), the electric field has a stabilizing effect, and the maximum growth rate, the most unstable wavenumber, and the cutoff wavenumber all decrease as $E_b$ increases. When $E_b$ is sufficiently large, the instability dominated by the Marangoni mode can be completely suppressed, causing the corresponding mode to become stable. In addition, \autoref{fig3} shows that the electric field exerts a pronounced destabilizing or stabilizing influence on the dominant mode, whereas its effect on the subdominant mode is comparatively weak. This behaviour helps explain the gradual separation of the two growth-rate branches at sufficiently large $E_b$ in \autoref{fig2}($b$). Although each growth-rate branch may contain different segments dominated by the RP and Marangoni modes, respectively, the electric field primarily modifies the dominant branch with the larger growth rate. Consequently, the original crossing and connection between the two modes are altered, and the two growth-rate branches eventually become separated.
\begin{figure}
    \centering
    \begin{subfigure}[b]{0.45\textwidth}
        \centering
        \includegraphics[width=\textwidth, trim={0pt 0pt 0pt 0pt}, clip]{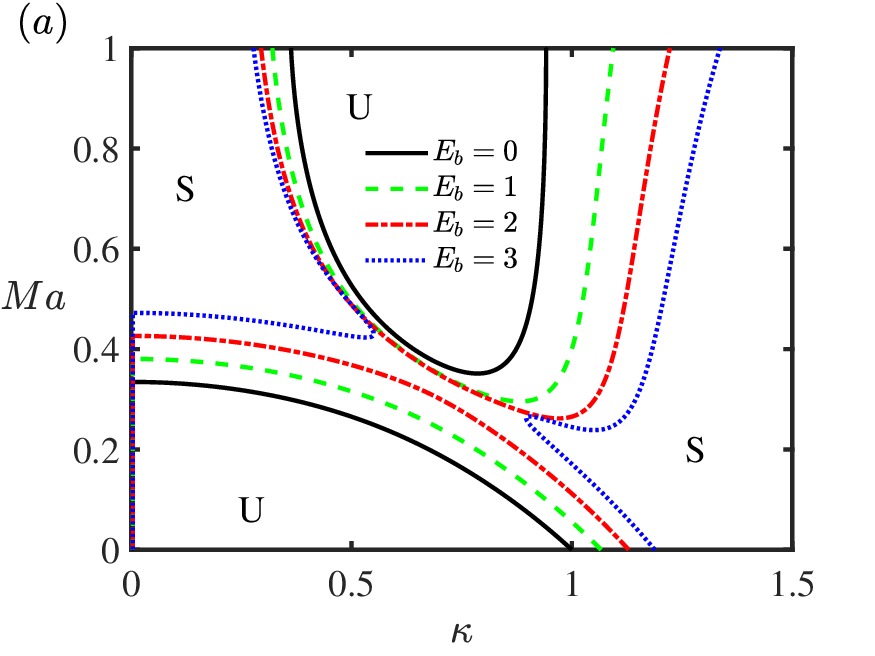}
    \end{subfigure}
    \hspace{0.02\textwidth}
    \begin{subfigure}[b]{0.45\textwidth}
        \centering
        \includegraphics[width=\textwidth, trim={0pt 0pt 0pt 0pt}, clip]{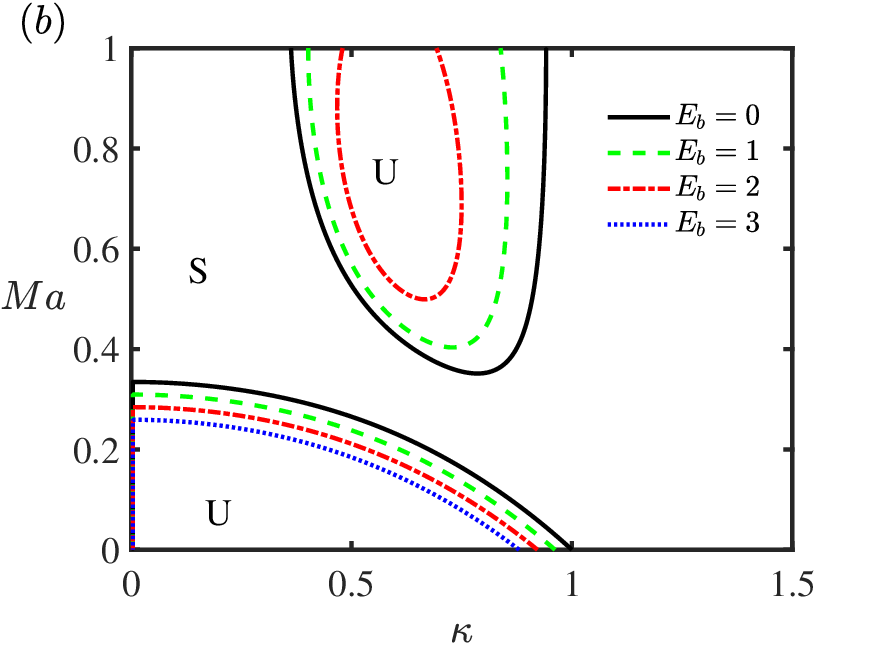}
    \end{subfigure}
    \caption{\raggedright Variation of the neutral stability boundaries with the electric Weber number $E_b$ for two outer-electrode positions: $(a)$ $\beta=e^{0.9}$ and $(b)$ $\beta=e^{1.1}$.}
    \label{fig4}
\end{figure}

To further reveal the coupling between the electric field and the Marangoni effect, \autoref{fig4} presents the neutral stability boundaries in the $Ma$-$\kappa$ plane at different values of $E_b$. \autoref{fig4}($a$) corresponds to the case $\beta<e$. In the absence of an electric field, the $Ma$-$\kappa$ plane is divided into three regions: an unstable region dominated by the RP mode at low $Ma$, an unstable region dominated by the Marangoni mode at high $Ma$, and a stable region lying between and outside the two unstable regions. The symbols `U' and `S' denote the unstable and stable regions, respectively. As $E_b$ increases, the two unstable regions expand and gradually approach one another, eventually becoming fully connected at $E_b=3$. By contrast, when $\beta>e$, as shown in \autoref{fig4}($b$), the radial electric field compresses both unstable regions. The RP-dominated and Marangoni-dominated unstable regions progressively shrink as $E_b$ increases, and the Marangoni-dominated unstable region is completely eliminated at $E_b=3$. This result suggests that, compared with the approach adopted by \citet{Ding2014}, in which a strong electric field is applied to suppress the interfacial mode directly, enhancing the Marangoni effect while applying a weaker electric field may provide a more practical route to film stabilization with a lower required electric-field strength. It is also noteworthy that, over certain ranges of wavenumber in \autoref{fig4}($a$), the electric field expands the RP-unstable region much more strongly than it affects the Marangoni-unstable region. The merging of the two unstable regions therefore occurs primarily through the extension of the RP-unstable region towards higher $Ma$, causing it to penetrate the region originally dominated by the Marangoni instability, rather than through a comparable expansion of both regions towards one another.
\begin{figure}
    \centering
    \begin{subfigure}[b]{0.45\textwidth}
        \centering
        \includegraphics[width=\textwidth, trim={0pt 0pt 0pt 0pt}, clip]{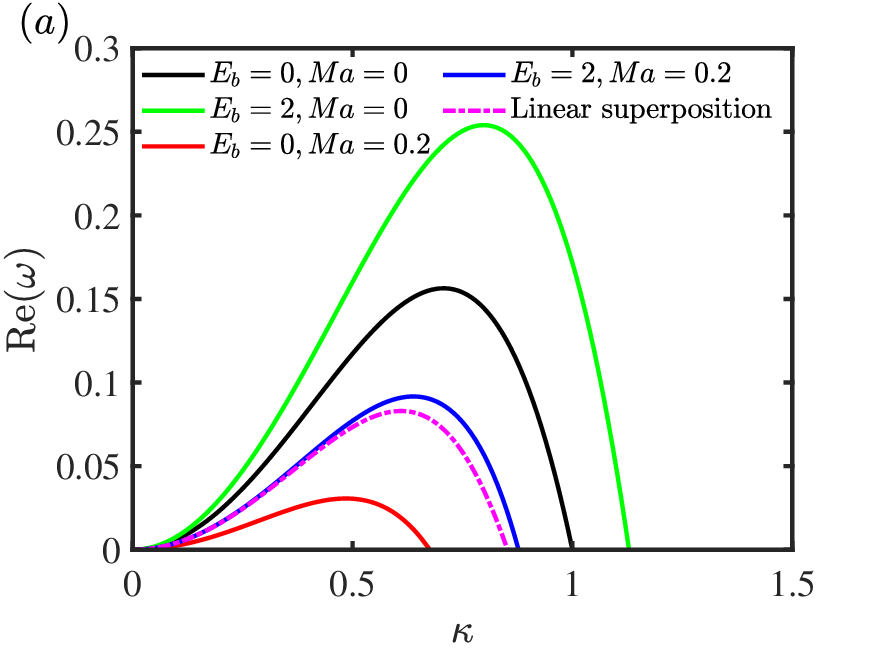}
        \label{fig5a}
    \end{subfigure}
    \hspace{0.02\textwidth}
    \begin{subfigure}[b]{0.45\textwidth}
        \centering
        \includegraphics[width=\textwidth, trim={0pt 0pt 0pt 0pt}, clip]{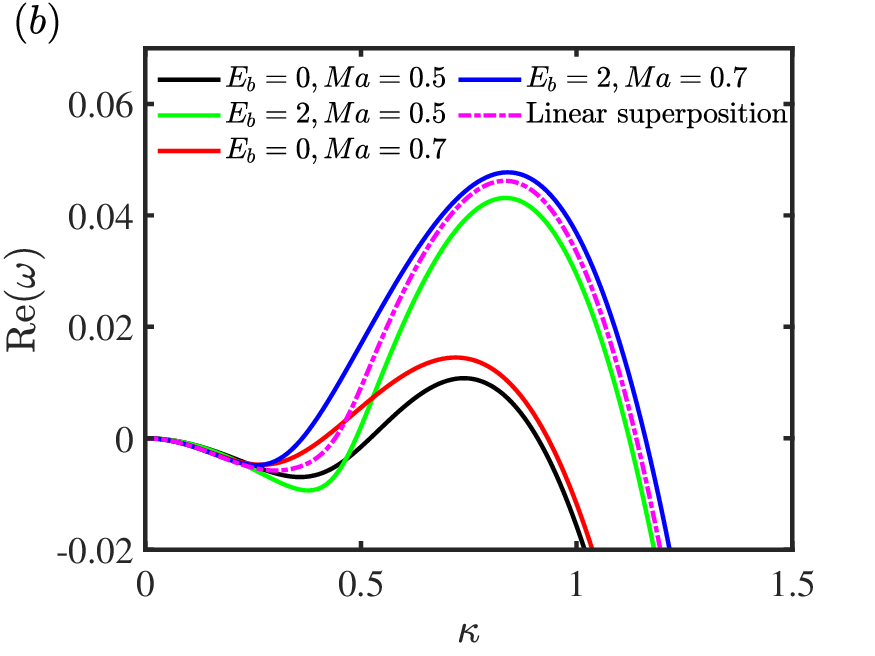}
        \label{fig5b}
    \end{subfigure}
    \caption{\raggedright Non-additive coupling between the electric field and Marangoni effects: $(a)$ the RP mode and $(b)$ the Marangoni mode. The magenta dash-dotted lines represent the predictions based on linear superposition. }
    \label{fig5}
\end{figure}

Motivated by the observation in \autoref{fig4}($a$) that the electric field promotes the two unstable modes to markedly different extents over certain ranges of wavenumber, we further examine whether the effects of the electric field and the Marangoni mechanism on film stability can be described by linear superposition. The results are presented in \autoref{fig5}, where \autoref{fig5}($a$) corresponds to the regime dominated by the RP mode, whereas \autoref{fig5}($b$) corresponds to that dominated by the Marangoni mode. A reference case is first selected, after which the electric field and the Marangoni number are varied separately. The corresponding changes in growth rate relative to the reference state are then added to the reference growth rate to obtain the prediction based on linear superposition. This prediction is subsequently compared with the actual growth rate obtained when the electric field and the Marangoni effect act simultaneously. As shown in \autoref{fig5}, regardless of whether the RP mode or the Marangoni mode is dominant, the growth rate predicted by linear superposition is lower than the actual growth rate over certain ranges of wavenumber. This result demonstrates that the electric field and the Marangoni effect do not influence film stability independently, but instead exhibit a pronounced non-additive coupling. A straightforward expectation would be that, although the electric field promotes both the RP and Marangoni instabilities, the associated enhancement of Marangoni convection could further suppress the RP instability, causing the actual growth rate to fall below the prediction based on linear superposition. The opposite trend observed in the calculations therefore indicates that the coupling between the electric field and surfactant transport introduces an additional destabilizing contribution, whose underlying physical mechanism requires further examination.

\section{Nonlinear dynamics and coupling mechanisms}
\label{Nonlinear dynamics and coupling mechanisms}
To address the questions left unresolved by the linear stability analysis and further elucidate the coupling mechanism between the electric field and the Marangoni effect, traveling wave solutions are employed to analyse the nonlinear dynamics of the system, following the approaches of \citet{Ding2014,Gao2026}. Before seeking the traveling wave solutions, the established one-dimensional model is first reformulated in the following conservative form:
\begin{equation}
    s_t+(2q)_z=0,
    \label{eq:conservation_s}
\end{equation}
\begin{equation}
    \left(\sqrt{s}\Gamma\right)_t
    +\left(\sqrt{s}w\Gamma\right)_z
    =
    \frac{1}{Pe}
    \left[\sqrt{s}\Gamma_z\right]_z,
    \label{eq:conservation_Gamma}
\end{equation}
where $s=(\alpha+h)^2$. It can be readily verified that the above equations are conservative under periodic boundary conditions. The remaining variables are rewritten as :
\begin{equation}
q={}-\frac{p_z-1}{4}
\left[
s^2\ln\left(\frac{\sqrt{s}}{\alpha}\right)
-\frac{\alpha^4-4\alpha^2s+3s^2}{4}
\right]
+\frac{\gamma_z}{2}
\left[
s^{3/2}\ln\left(\frac{\sqrt{s}}{\alpha}\right)
-\frac{\sqrt{s}\left(s-\alpha^2\right)}{2}
\right],
\label{eq:q_conservative}
\end{equation}
\begin{equation}
p=\frac{E_b}{2s\left[\ln\left(\sqrt{s}\right)-\ln\beta\right]^2}+\gamma\left[
\frac{1}{\sqrt{s}} -\varepsilon^2\left(\sqrt{s}\right)_{zz}\right],
\label{eq:p_conservative}
\end{equation}
and
\begin{equation}
w=\frac{p_z-1}{4}\left(s-\alpha^2\right)+
\left[\sqrt{s}\gamma_z-\frac{p_z-1}{2}s\right]
\ln\left(\frac{\sqrt{s}}{\alpha}\right),
\label{eq:w_conservative}
\end{equation}
where $\gamma=1+Ma\ln(1-\Gamma)$. Prior to the nonlinear evolution, a small harmonic perturbation is imposed on the interface of the base flow at the initial time:
\begin{equation}
s(z,0)=\left[1+0.1\sin\left( kz\right)\right]^2,
\label{eq:initial_perturbation}
\end{equation}
where $k$ denotes the selected disturbance wavenumber and is related to the computational-domain length $L$ through $k=2\pi/L$. The governing equations are discretized in space using a Fourier spectral method, which automatically satisfies the periodic boundary conditions. Time integration is performed using an implicit multistep scheme, while the resulting nonlinear algebraic equations are solved through Newton iteration. A total of 256 Fourier collocation points are employed, and the relative error tolerance is set to $10^{-10}$. Further details of the numerical procedure can be found in \citet{Gao2026}. Nonlinear simulations are conducted because traveling wave branches frequently bifurcate near the cutoff wavenumber and may contain spurious solutions that cannot be dynamically realized. Comparison with the nonlinear evolution therefore provides a means of determining whether a traveling wave solution can develop from a finite-amplitude disturbance. Alternatively, the stability of the traveling wave solutions can be analysed directly to identify the stability of the different bifurcation branches \cite{Camassa2016}. We next introduce the traveling wave coordinate $\xi=z-ct$, where $c$ denotes the wave speed. Under this coordinate transformation, integrating \eqref{eq:conservation_s} and \eqref{eq:conservation_Gamma} once with respect to $\xi$ yields the following forms:
\begin{equation}
    -c s+2q=q_0,
\label{TW: s}
\end{equation}
\begin{equation}
-c\sqrt{s}\,\Gamma+\sqrt{s}\,w\Gamma
=\frac{1}{Pe}\sqrt{s}\,\Gamma_{\xi}+Q_0,
\label{TW: Gamma}
\end{equation}
where $q_0$ and $Q_0$ are integration constants. The unknowns in \eqref{TW: s} and \eqref{TW: Gamma} comprise the two functions $s$ and $\sqrt{s}\Gamma$, together with the wave speed $c$ and the integration constants $q_0$ and $Q_0$. Starting from appropriate initial guesses, the corresponding traveling wave solutions can be obtained by solving this nonlinear system. To close the system and remove the translational indeterminacy of the traveling wave solutions, two mass-conservation constraints and one phase condition must also be imposed, thereby yielding a unique converged solution. The additional conditions are given by :
\begin{equation}
    \frac{1}{L}\int_{0}^{L}s\,\mathrm{d}\xi=1,
\label{TW: s conservation}
\end{equation}

\begin{equation}
\frac{1}
{L}\int_{0}^{L}\left(\sqrt{s}\,\Gamma\right)\mathrm{d}\xi
=\bar{\Gamma},
\label{TW: Gamma conservation}
\end{equation}

\begin{equation}
    s(0)=1.
\label{TW: phase condition}
\end{equation}
It should be further noted that the computation of the traveling wave solutions is initiated in the vicinity of the cutoff wavenumber. The initial guess for the interfacial variable $s$ is prescribed in the harmonic form given by \eqref{eq:initial_perturbation}, while the initial guess for $\Gamma$ is taken as the uniform initial interfacial concentration. The initial estimate of the wave speed $c$ is determined from the phase speed in the long-wave limit. Specifically, twice the base-state interfacial velocity is used when the RP mode is dominant, whereas the base-state interfacial velocity is adopted when the Marangoni mode is dominant. The initial values of the integration constants $q_0$ and $Q_0$ are then evaluated from \eqref{TW: s} and \eqref{TW: Gamma}, respectively. Once an initial traveling wave solution has been obtained near the cutoff wavenumber, the complete solution branch is traced by taking the wavenumber as the continuation parameter and applying a continuation method.

\begin{figure}
    \centering
    \begin{subfigure}[b]{0.45\textwidth}
        \centering
        \includegraphics[width=\textwidth,
      trim={0pt 0pt 0pt 0pt}, clip]{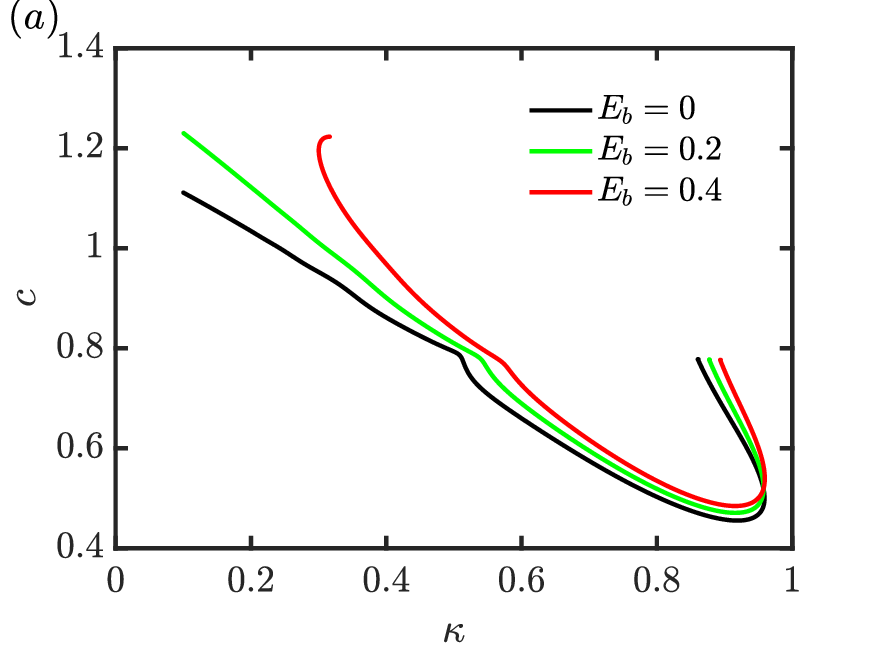}
    \end{subfigure}
    \hspace{0.02\textwidth}
    \begin{subfigure}[b]{0.45\textwidth}
        \centering
        \includegraphics[width=\textwidth,
        trim={0pt 0pt 0pt 0pt},clip]{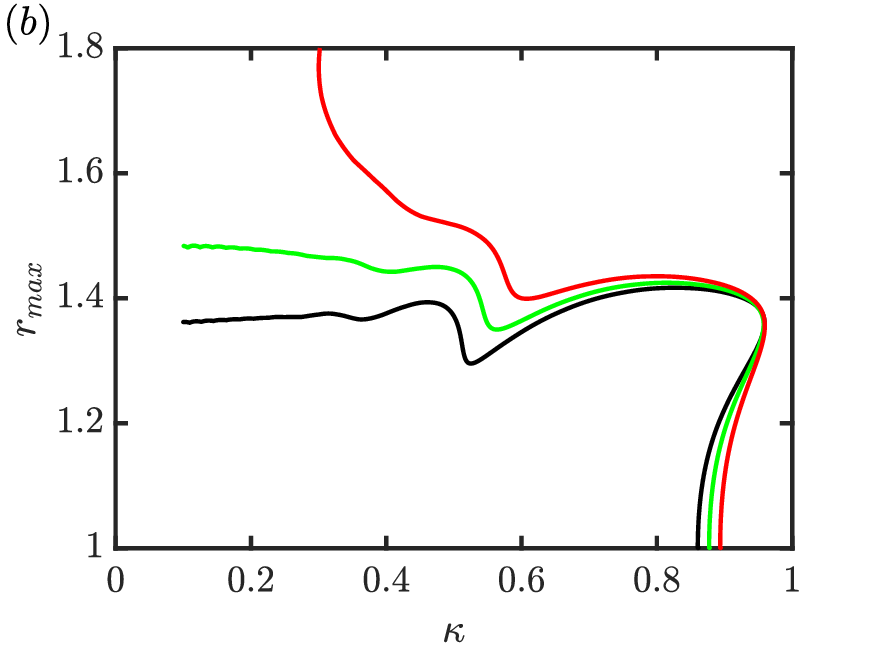}
    \end{subfigure}
    \vspace{0.02\textwidth}
    \begin{subfigure}[b]{0.45\textwidth}
        \centering
        \includegraphics[width=\textwidth,
            trim={0pt 0pt 0pt 0pt},clip]{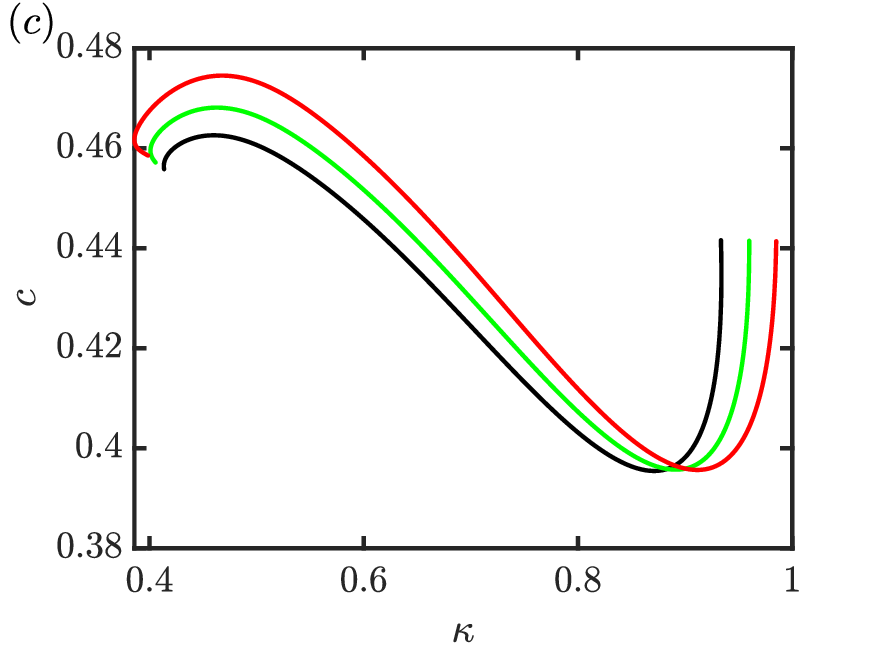}
    \end{subfigure}
    \hspace{0.02\textwidth}
    \begin{subfigure}[b]{0.45\textwidth}
        \centering
        \includegraphics[width=\textwidth,trim={0pt 0pt 0pt 0pt},clip]{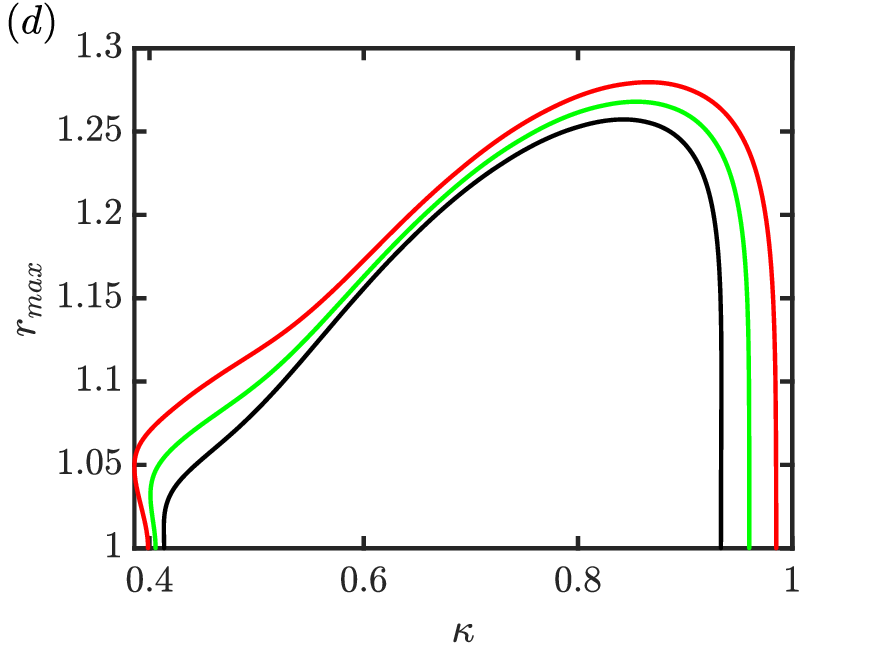}
    \end{subfigure}
    \caption{\raggedright traveling wave branches for different electric Weber numbers $E_b$ at $\beta=e^{0.9}<e$: $(a,b)$ $Ma=0.1$ and $(c,d)$ $Ma=0.7$. The wave speed $c$ and the maximum interfacial radius $r_{\max}=\max_{\xi}(\alpha+h)$ are plotted against the wavenumber $\kappa$ in the left and right panels, respectively.}
    \label{fig6}
\end{figure}

To examine separately the influence of the electric field on traveling wave branches in the regimes dominated by the RP mode and the Marangoni mode, $Ma=0.1$ and $Ma=0.7$ are selected as representative conditions. The former lies within the parameter region dominated by the RP mode, whereas the latter corresponds to the region dominated by the Marangoni mode. Only the case $\beta<e$ is considered here, for which the radial electric field is destabilizing and is therefore more suitable for investigating the competition and coupling between the electric and Marangoni effects. The case $\beta>e$, in which the electric field has a stabilizing effect, is not considered in the subsequent discussion.

As shown in \autoref{fig6}($a,b$), when $Ma=0.1$, the nontrivial traveling wave branches bifurcate from the vicinity of the cutoff wavenumber. As $E_b$ increases, both the wave speed $c$ and the maximum interfacial radius $r_{\max}$ generally increase at a given wavenumber, indicating that the electric field substantially alters the propagation speed and the amplitude of interfacial deformation. In the present calculations, the maximum value of $E_b$ is limited to $0.4$, which is considerably smaller than the range considered in the linear stability analysis. This is because, even at $E_b=0.4$, the traveling wave branch undergoes a pronounced secondary turning at small wavenumbers, and the continuation is therefore terminated at this point. Correspondingly, $r_{\max}$ in \autoref{fig6}($b$) increases rapidly at small wavenumbers and progressively departs from the regular traveling wave branches obtained under weaker electric fields. Previous studies have shown that such branches may continue towards a touchdown singularity, at which the interface comes into contact with the outer electrode \citep{Ding2014}. It should be emphasized that the solutions beyond the turning point cannot be classified as nonphysical solely from the shape of the branch, because some may remain mathematically valid while being dynamically unstable. Simulations of the nonlinear temporal evolution can be used to determine whether these traveling wave solutions can evolve from disturbances of finite amplitude and thereby identify the stable traveling waves that are dynamically realizable. When $Ma=0.7$, the traveling wave branches exhibit characteristics distinct from those found in the region dominated by the RP mode, as shown in \autoref{fig6}($c,d$). The wave speed $c$ varies nonmonotonically with the wavenumber, rather than increasing continuously as the wavenumber decreases. The maximum interfacial radius $r_{\max}$ also first increases and then decreases with the wavenumber, showing an overall trend similar to that of the growth rate of the Marangoni mode obtained from the linear stability analysis. As $E_b$ increases, both the wave speed and the maximum interfacial radius generally increase over the wavenumber range considered. The electric field therefore also enhances the response of the traveling waves in the region dominated by the Marangoni mode, although the corresponding branch structure differs markedly from that found in the region dominated by the RP mode. These results further show that, at small wavenumbers, traveling wave branches subjected to an electric field are more likely to exhibit turning behaviour and a rapid increase in interfacial amplitude. Since the primary objective of the present study is to compare the effects of electric forcing and Marangoni stresses on traveling wave states, the subsequent analysis is performed at $\kappa=1/\sqrt{2}$, which is the most unstable wavenumber of the reference state without surfactant or electric forcing. On this basis, $Ma$ and $E_b$ are employed separately as continuation parameters to examine the dependence of the traveling wave solutions on these two control parameters.

\subsection{Competition governing the surfactant effect}
\label{Competition governing the surfactant effect}
\begin{figure}
    \centering
    \begin{subfigure}[b]{0.45\textwidth}
        \centering
        \includegraphics[width=\textwidth,
      trim={0pt 0pt 0pt 0pt}, clip]{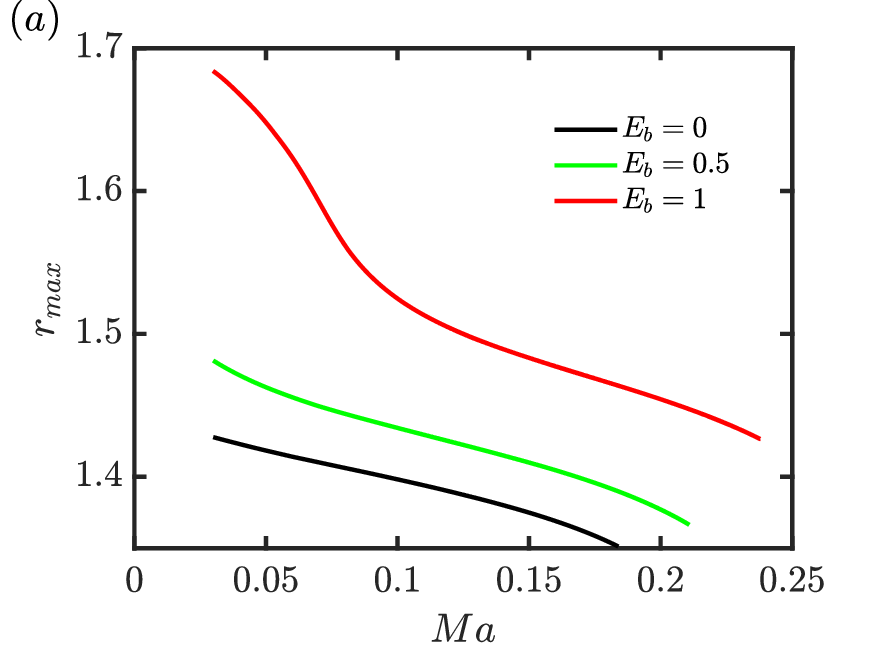}
    \end{subfigure}
    \hspace{0.02\textwidth}
    \begin{subfigure}[b]{0.45\textwidth}
        \centering
        \includegraphics[width=\textwidth,
        trim={0pt 0pt 0pt 0pt},clip]{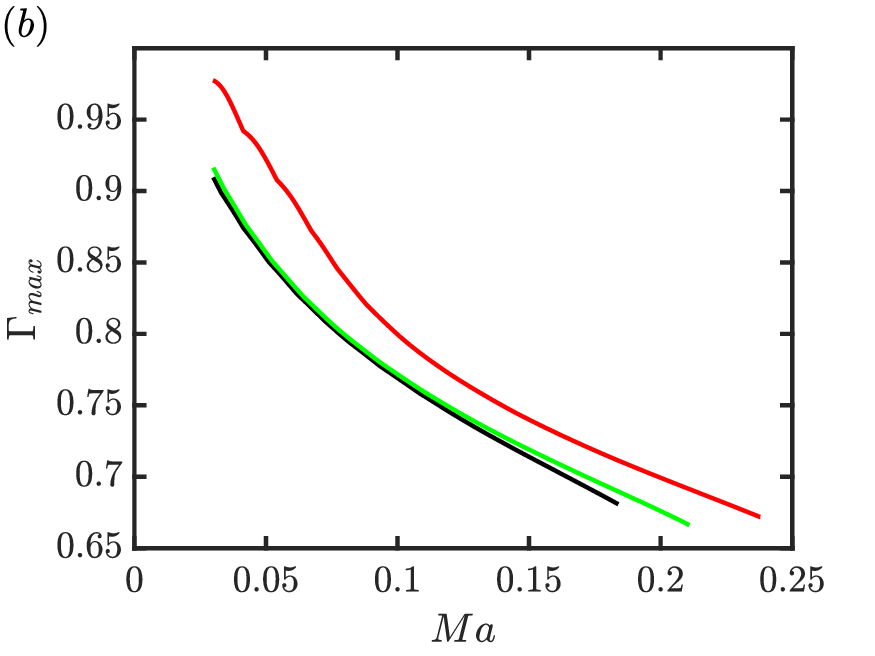}
    \end{subfigure}
    \vspace{0.02\textwidth}
    \begin{subfigure}[b]{0.45\textwidth}
        \centering
        \includegraphics[width=\textwidth,
            trim={0pt 0pt 0pt 0pt},clip]{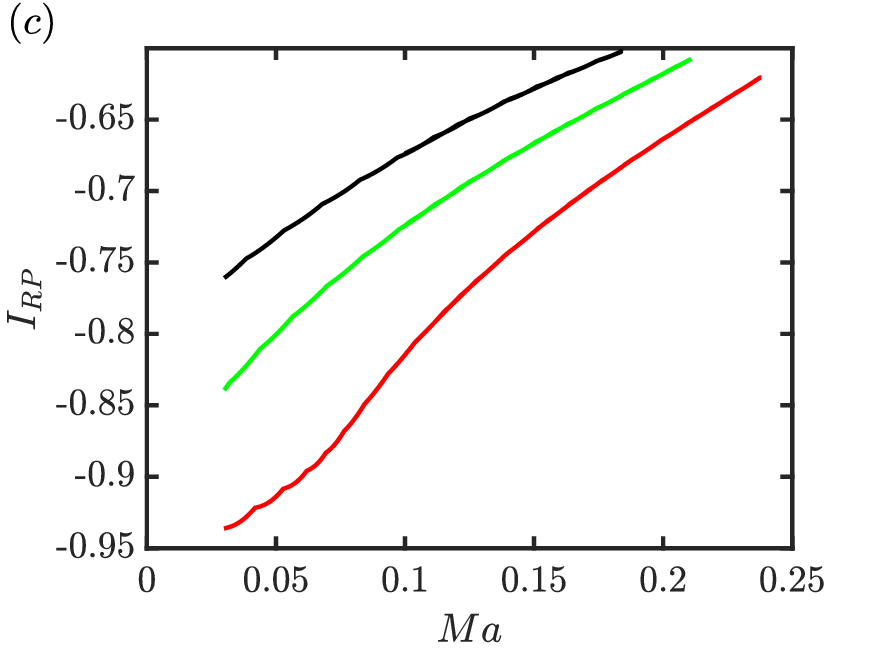}
    \end{subfigure}
    \hspace{0.02\textwidth}
    \begin{subfigure}[b]{0.45\textwidth}
        \centering
        \includegraphics[width=\textwidth,trim={0pt 0pt 0pt 0pt},clip]{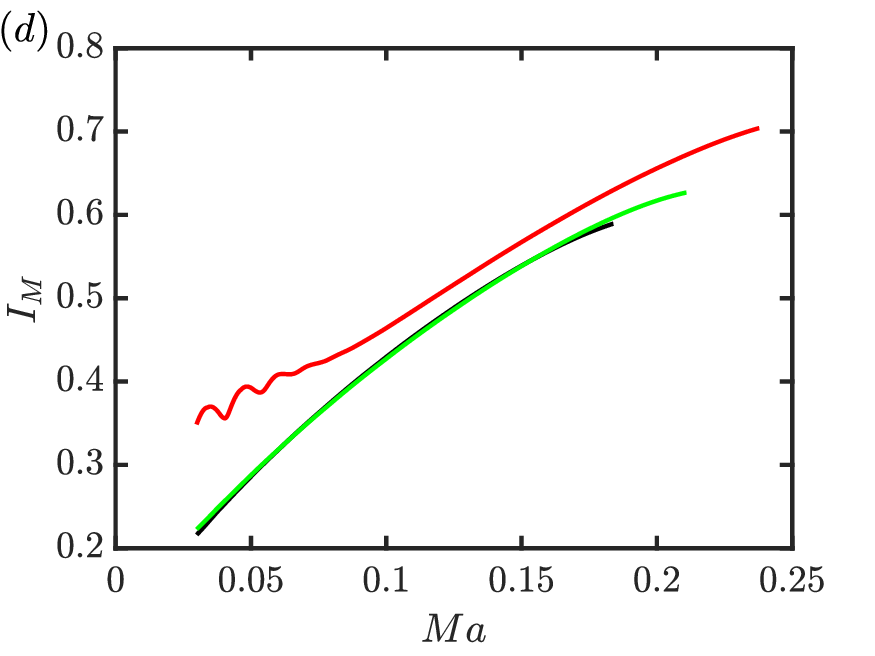}
    \end{subfigure}
    \caption{\raggedright Traveling wave branches obtained by continuation in $Ma$ from the reference state $Ma=0.1$ at $\kappa=1/\sqrt{2}$ and $\beta=e^{0.9}<e$ for different values of $E_b$: $(a)$ maximum interfacial radius $r_{\max}$, $(b)$ maximum interfacial surfactant concentration $\Gamma_{\max}$, $(c)$ intensity of the $c_w$, and $(d)$ intensity of Marangoni convection $I_{\mathrm{M}}$.}
    \label{fig7}
\end{figure}

\citet{Gao2026} proposed a competition mechanism between relative interfacial motion and Marangoni convection. The relative interfacial velocity is defined as $c_w=c-w$, whereas Marangoni convection is characterized by the surface tension gradient $\gamma_{\xi}$. Since both $c_w$ and $\gamma_{\xi}$ vary with the spatial coordinate, the following quantities are introduced to quantify their respective intensities:
\begin{equation}
    I_{\mathrm{RP}}
    =
    -\max_{\xi}\left|c_w\right|,
    \qquad
    I_{\mathrm{M}}
    =
    \frac{1}{L}
    \int_{0}^{L}
    \max\left(\gamma_{\xi},0\right)
    \,\mathrm{d}\xi .
    \label{eq:mode_intensity}
\end{equation}

The quantity $I_{RP}$ is defined as the negative value of the maximum magnitude of the relative interfacial velocity because, in the traveling wave coordinate system, the liquid moves relative to the wave in the direction opposite to the actual direction of wave propagation. The negative sign of $I_{\mathrm{RP}}$ therefore indicates the direction of the relative interfacial motion, whereas its intensity should be evaluated using $\left|I_{RP}\right|$. For Marangoni convection, $\gamma_{\xi}$ satisfies periodic boundary conditions, and its signed integral over one period is therefore zero. The intensity $I_{M}$ is consequently defined using the spatial average of the positive part of $\gamma_{\xi}$. This definition provides a more direct measure of the ability of the Marangoni stress to transport liquid along the interface. \autoref{fig7} presents the results obtained by continuation in $Ma$ from the reference solution at $Ma=0.1$. As shown in \autoref{fig7}($a-b$), both the maximum interfacial radius $r_{\max}$ and the maximum surfactant concentration $\Gamma_{\max}$ decrease as $Ma$ increases. It should be noted that the lower limit of this traveling wave branch does not extend to $Ma=0$. As $Ma$ approaches zero, the ability of the Marangoni stress to regulate the interfacial surfactant distribution becomes progressively weaker. The surfactant is then transported by the base flow and the relative interfacial motion towards the interfacial bulge, resulting in a continuous increase in the local surfactant concentration. When the local concentration approaches the saturation limit permitted by the Langmuir equation of state, the equation becomes singular and the traveling wave solution can no longer be continued. As shown in \autoref{fig7}($c-d$), the intensity of the relative interfacial motion, $\left|I_{RP}\right|$, decreases with increasing $Ma$, whereas the intensity of Marangoni convection, $I_{M}$, increases. This trend is consistent with the results of \citet{Gao2026}. In the parameter region where the RP mode is dominant, the surfactant acts through two principal mechanisms. First, it reduces the interfacial surface tension and thereby weakens the capillary effect that drives the RP instability. Second, Marangoni convection acts in the direction opposite to the transport induced by the relative interfacial motion and therefore impedes the further accumulation of liquid and surfactant at the interfacial bulge. Consequently, as $I_{M}$ increases, both $\left|I_{RP}\right|$ and $r_{\max}$ decrease, indicating that Marangoni convection mainly suppresses the RP instability in this region. To compare the two modes further, the effect of the electric field is temporarily excluded, and the continuation results obtained from the reference solution at $Ma=0.7$ in \autoref{fig8} are examined. As shown in \autoref{fig8}($a,d$), in the unstable region where the Marangoni mode is dominant, the increase in the amplitude of the interfacial deformation occurs together with the emergence of Marangoni convection and varies with $I_{M}$. This correspondence indicates that the unstable branch is triggered by Marangoni convection and that its strength is governed primarily by the intensity of that convection. Meanwhile, as shown in \autoref{fig8}($b$), the interface becomes more rigid at larger values of $Ma$, which suppresses the local accumulation of surfactant at the interfacial bulge. The local concentration therefore does not continue to increase towards the limit of the equation of state, as occurs at smaller values of $Ma$. \autoref{fig8}($c$) also reveals a feature that differs from the behaviour observed in the region where the RP mode is dominant. Within the smaller range of $Ma$ in which the Marangoni mode begins to emerge, the intensity of the relative interfacial motion, $\left|I_{RP}\right|$, and the intensity of Marangoni convection, $I_{M}$, increase simultaneously. This occurs because the formation of the Marangoni stress depends on a gradient in the interfacial surfactant concentration, and this gradient must first be established through the transport of surfactant by the relative interfacial motion. The generation of Marangoni convection therefore depends on the relative interfacial motion, which also partly explains why the instability associated with the Marangoni mode is generally weaker than that associated with the RP mode. The essential mechanism through which surfactants affect the stability of the liquid film is the competition between Marangoni convection and relative interfacial motion.

When the electric field is introduced, the intensity of the relative interfacial motion, $\left|I_{RP}\right|$, increases substantially with $E_b$, whereas the intensity of Marangoni convection, $I_{M}$, changes only slightly at smaller values of $E_b$ and begins to increase noticeably only when the electric field becomes stronger. Similar trends are observed in the parameter regions where the RP mode and the Marangoni mode are dominant. The results in \autoref{fig7} and \autoref{fig8} therefore show that the electric field consistently tends to enhance the relative interfacial motion, whereas its influence on Marangoni convection remains comparatively weak. This behaviour persists even in the region where the Marangoni mode is dominant.

\begin{figure}
    \centering
    \begin{subfigure}[b]{0.45\textwidth}
        \centering
        \includegraphics[width=\textwidth,
      trim={0pt 0pt 0pt 0pt}, clip]{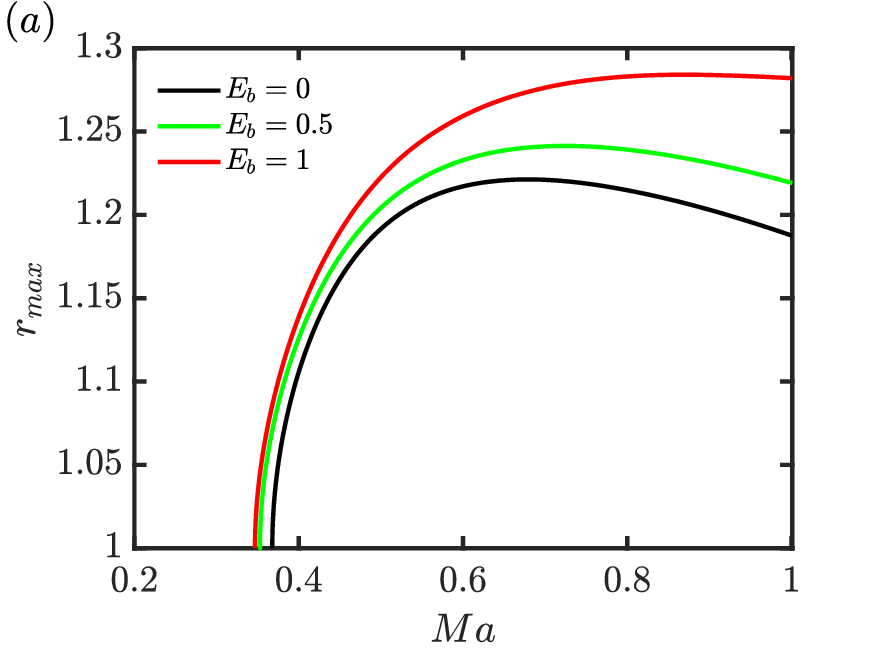}
    \end{subfigure}
    \hspace{0.02\textwidth}
    \begin{subfigure}[b]{0.45\textwidth}
        \centering
        \includegraphics[width=\textwidth,
        trim={0pt 0pt 0pt 0pt},clip]{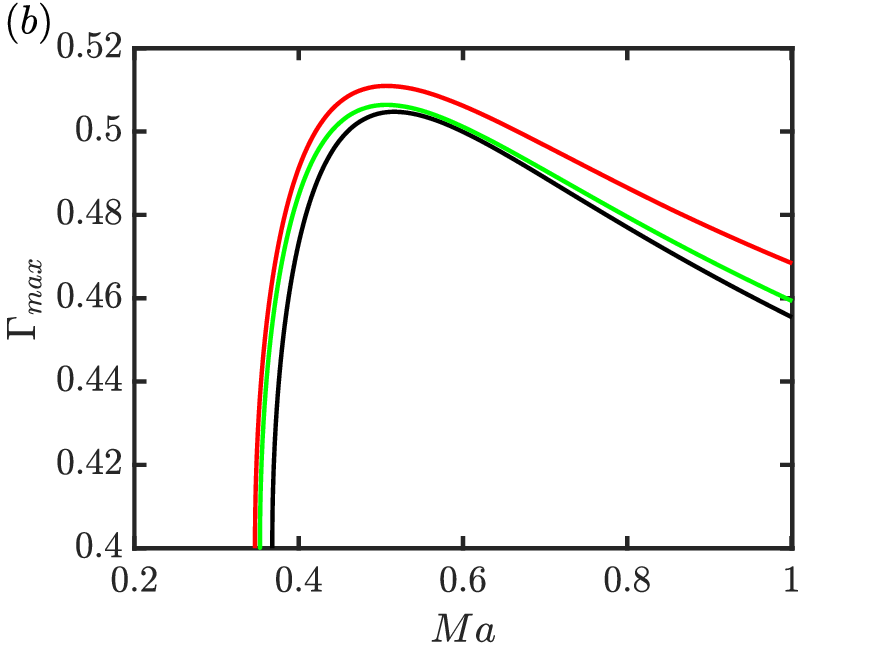}
    \end{subfigure}
    \vspace{0.02\textwidth}
    \begin{subfigure}[b]{0.45\textwidth}
        \centering
        \includegraphics[width=\textwidth,
            trim={0pt 0pt 0pt 0pt},clip]{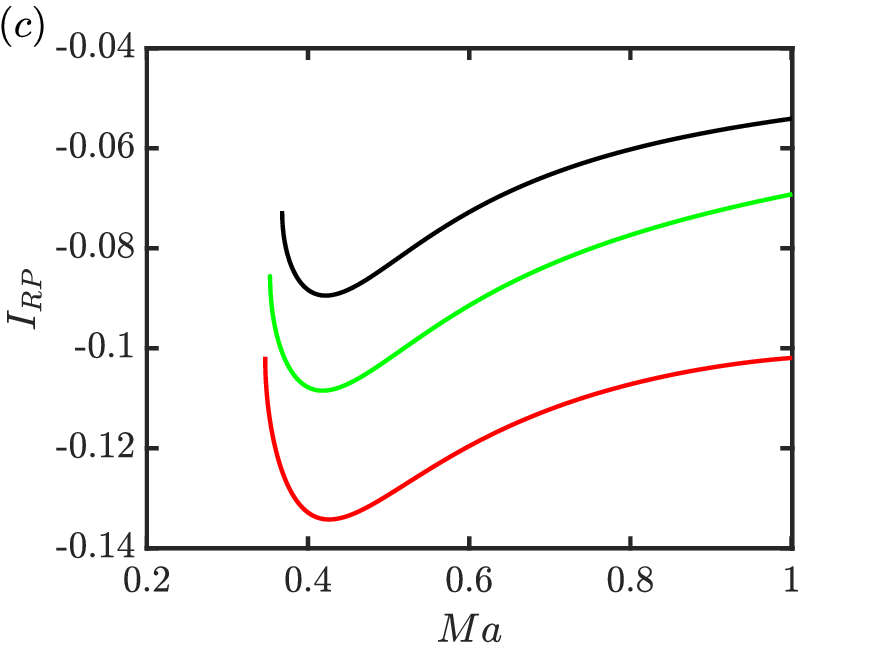}
    \end{subfigure}
    \hspace{0.02\textwidth}
    \begin{subfigure}[b]{0.45\textwidth}
        \centering
        \includegraphics[width=\textwidth,trim={0pt 0pt 0pt 0pt},clip]{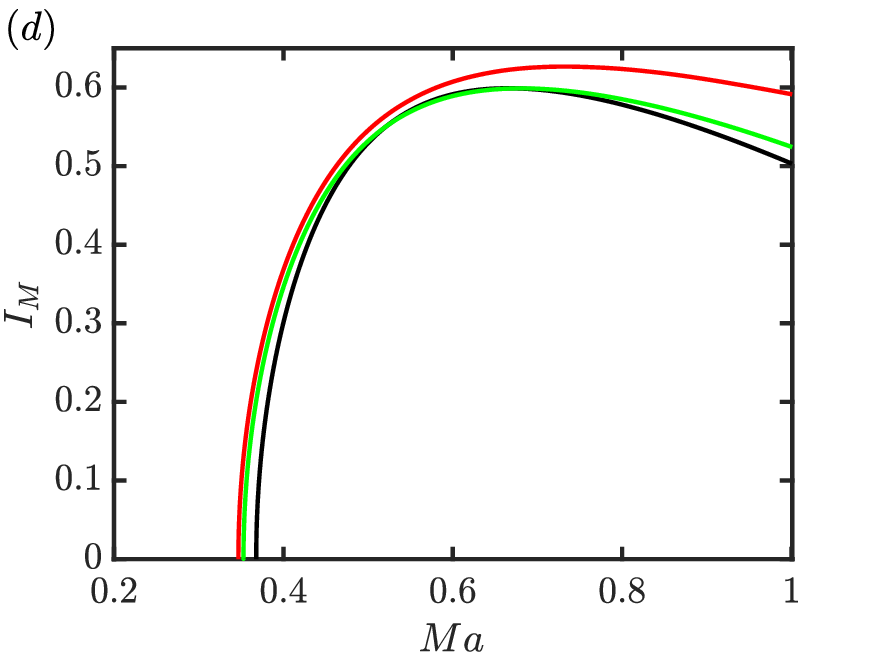}
    \end{subfigure}
    \caption{\raggedright Traveling wave branches obtained by continuation in $Ma$ from the reference solution at $Ma=0.7$, with $\kappa=1/\sqrt{2}$ and $\beta=e^{0.9}<e$, for different values of $E_b$: $(a)$ maximum interfacial radius $r_{\max}$, $(b)$ maximum interfacial surfactant concentration $\Gamma_{\max}$, $(c)$ intensity of the $c_w$ $I_{\mathrm{RP}}$, and $(d)$ intensity of Marangoni convection $I_{\mathrm{M}}$.}
    \label{fig8}
\end{figure}

\subsection{Modification of the competition by electric forcing} 
\label{Modification of the competition by electric forcing}

\begin{figure}
    \centering
    \begin{subfigure}[b]{0.45\textwidth}
        \centering
        \includegraphics[width=\textwidth,
      trim={0pt 0pt 0pt 0pt}, clip]{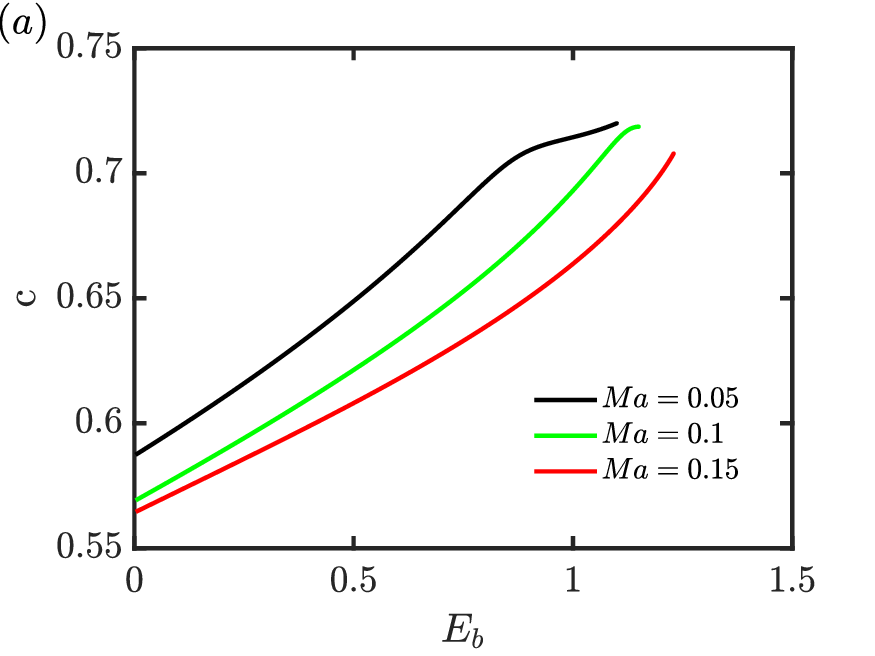}
    \end{subfigure}
    \hspace{0.02\textwidth}
    \begin{subfigure}[b]{0.45\textwidth}
        \centering
        \includegraphics[width=\textwidth,
        trim={0pt 0pt 0pt 0pt},clip]{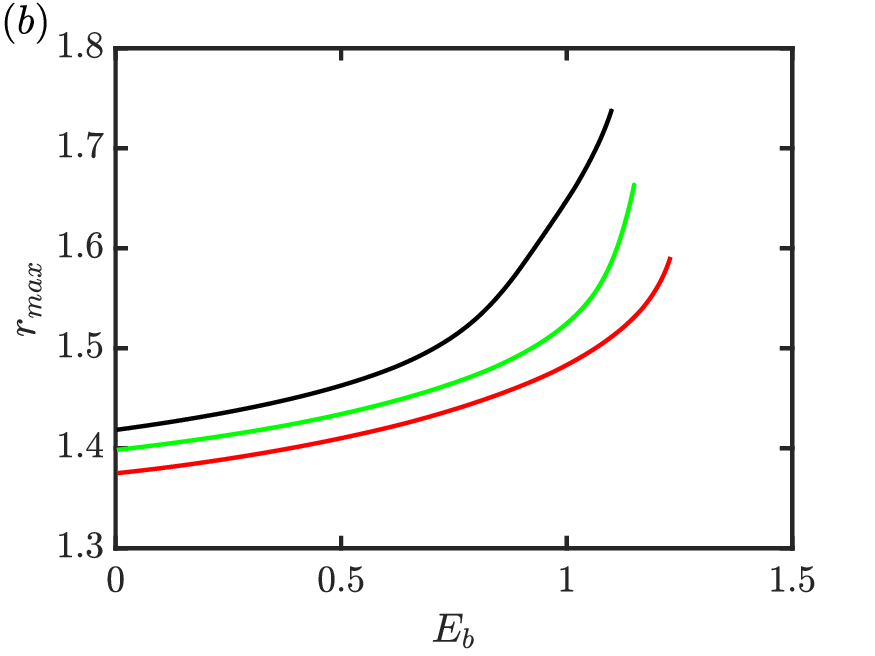}
    \end{subfigure}
    \vspace{0.02\textwidth}
    \begin{subfigure}[b]{0.45\textwidth}
        \centering
        \includegraphics[width=\textwidth,
            trim={0pt 0pt 0pt 0pt},clip]{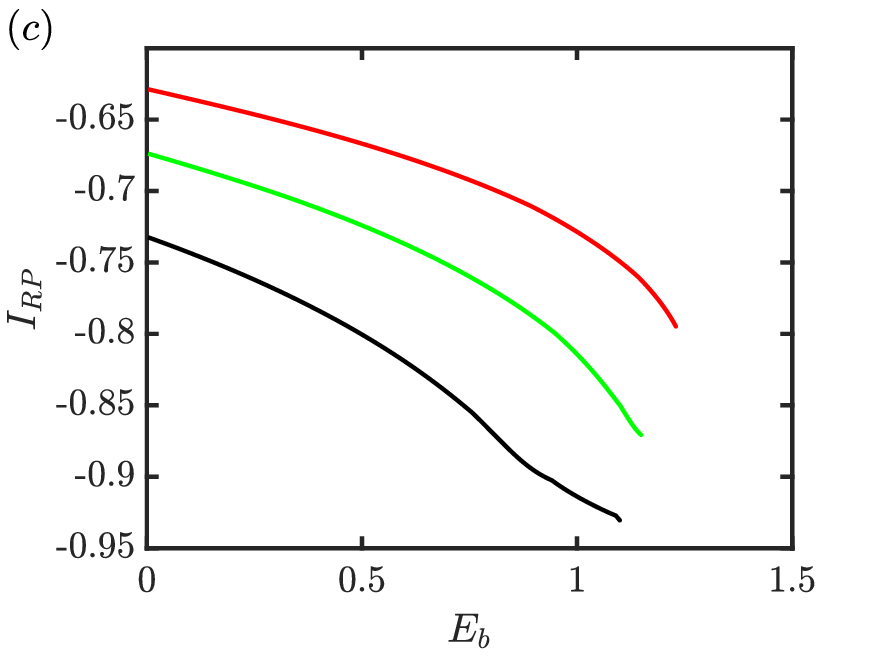}
    \end{subfigure}
    \hspace{0.02\textwidth}
    \begin{subfigure}[b]{0.45\textwidth}
        \centering
        \includegraphics[width=\textwidth,trim={0pt 0pt 0pt 0pt},clip]{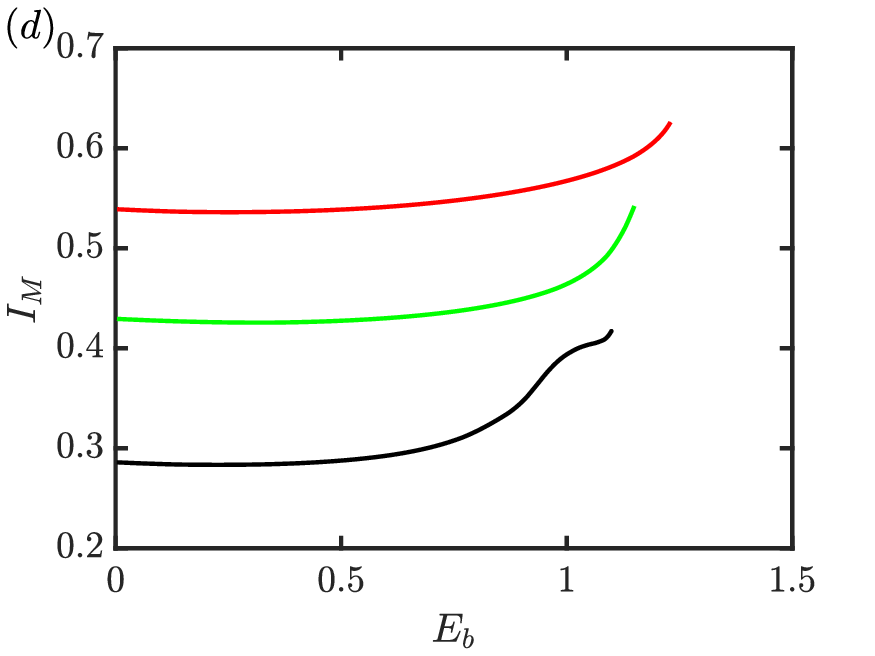}
    \end{subfigure}
    \caption{\raggedright Continuation results obtained by varying the electric Weber number $E_b$ in the region dominated by the RP mode for different values of $Ma$, with $\kappa=1/\sqrt{2}$: $(a)$ wave speed $c$, $(b)$ maximum interfacial radius $r_{\max}$, $(c)$ intensity of the relative interfacial motion $I_{RP}$, and $(d)$ intensity of Marangoni convection $I_{M}$.}
    \label{fig9}
\end{figure}

\begin{figure}
    \centering

    \begin{subfigure}[t]{0.32\textwidth}
        \centering
        \includegraphics[width=\textwidth]{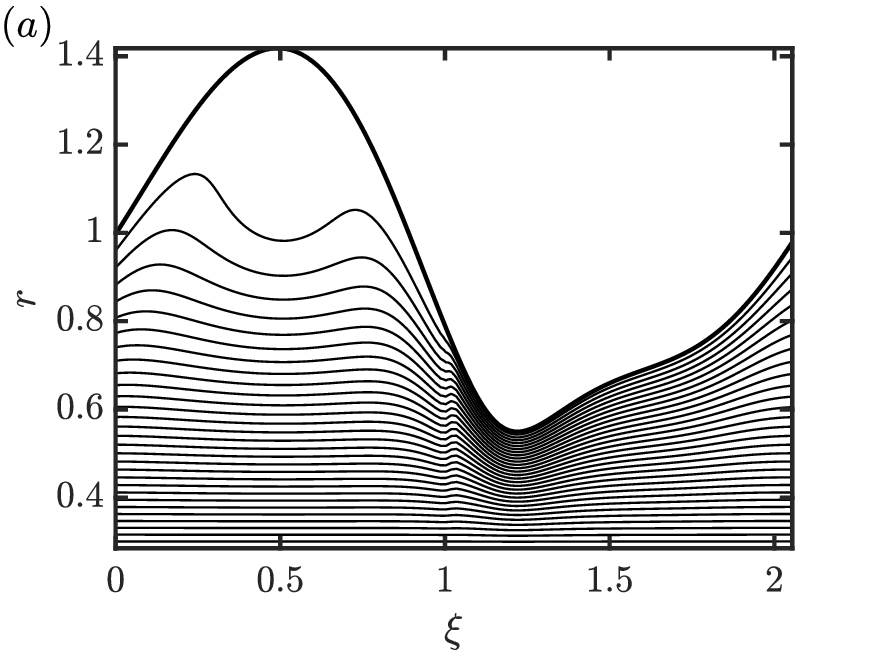}
    \end{subfigure}
    \hfill
    \begin{subfigure}[t]{0.32\textwidth}
        \centering
        \includegraphics[width=\textwidth]{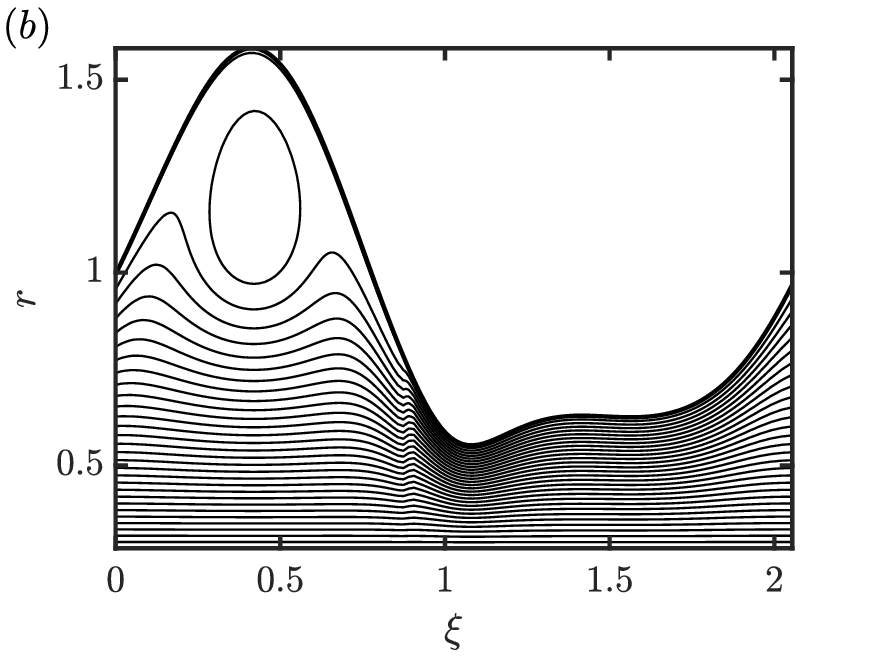}
    \end{subfigure}
    \hfill
    \begin{subfigure}[t]{0.32\textwidth}
        \centering
        \includegraphics[width=\textwidth]{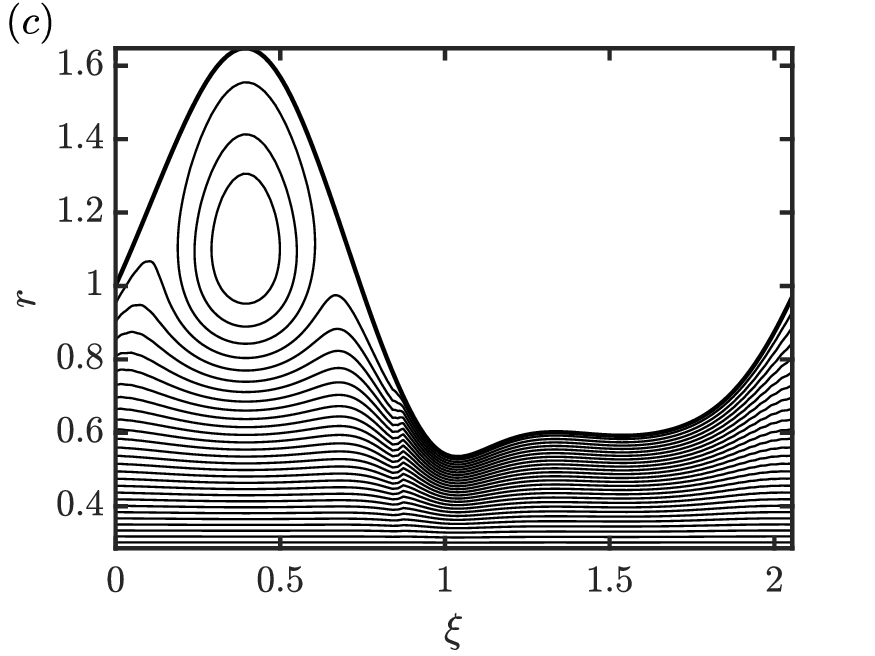}
    \end{subfigure}
    \caption{Streamline patterns of the traveling wave solutions at $Ma=0.05$ for different electric Weber numbers: $(a)$ $E_b=0$, $(b)$ $E_b=0.9$, and $(c)$ $E_b=1$.}
    \label{fig10}
\end{figure}

\begin{figure}
    \centering
    \begin{subfigure}[b]{0.45\textwidth}
        \centering
        \includegraphics[width=\textwidth, trim={0pt 0pt 0pt 0pt}, clip]{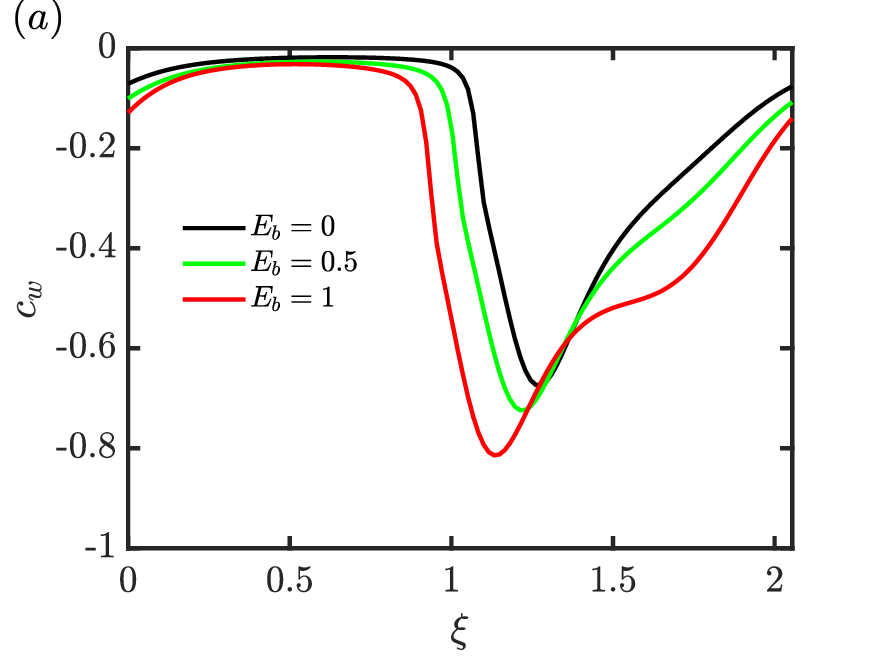}
    \end{subfigure}
    \hspace{0.02\textwidth}
    \begin{subfigure}[b]{0.45\textwidth}
        \centering
        \includegraphics[width=\textwidth, trim={0pt 0pt 0pt 0pt}, clip]{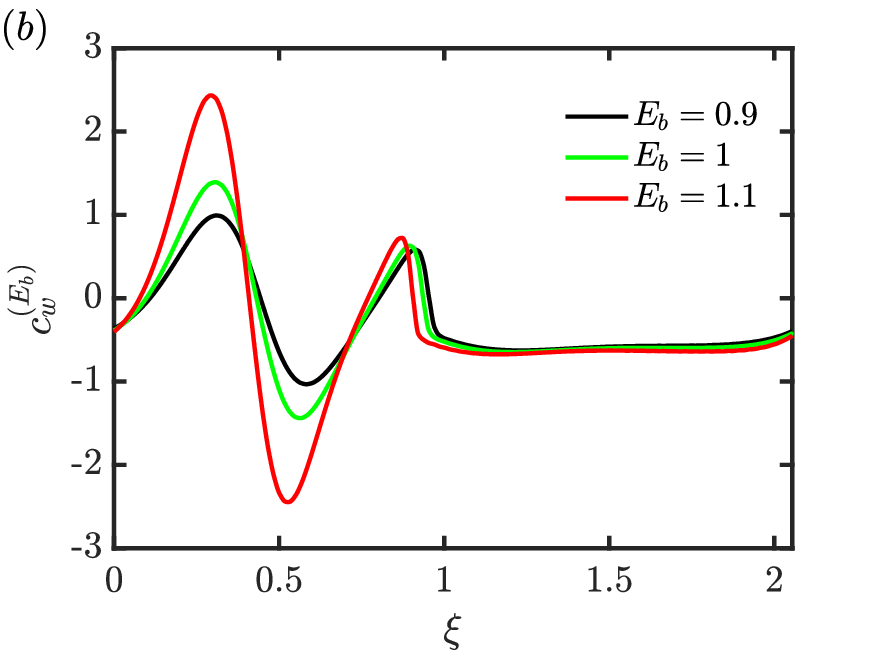}
    \end{subfigure}
    \caption{\raggedright Distributions of the interfacial relative velocity at $Ma=0.1$: $(a)$ the total relative velocity $c_w$ for different electric Weber numbers $E_b$, and $(b)$ the velocity contribution $c_w^{(E_b)}$ obtained by retaining only the effect of the electric stress gradient in $p_z$.} 
    \label{fig11}
\end{figure}

Since the electric field produces similar trends in the regions dominated by the RP mode and the Marangoni mode, the following discussion focuses on the region dominated by the RP mode. The aim is to clarify the physical mechanism associated with the electric field and to determine how this mechanism couples with the Marangoni effect. \autoref{fig9} shows how the principal characteristics of the traveling wave solutions vary with $E_b$ for different values of $Ma$. As $E_b$ increases, the wave speed $c$, the maximum interfacial radius $r_{\max}$, and the magnitude of the relative interfacial motion $\left|I_{RP}\right|$ all increase substantially. Moreover, the corresponding trends depend only weakly on $Ma$. By contrast, the intensity of Marangoni convection $I_M$ remains almost unchanged at smaller values of $E_b$ and increases appreciably only after the electric field becomes sufficiently strong. These results indicate that the electric field first enhances the relative interfacial motion, whereas the increase in Marangoni convection occurs only after the electric forcing exceeds a certain level. To examine how the electric field modifies the internal flow structure, \autoref{fig10} presents the streamline patterns at $Ma=0.05$ for different values of $E_b$. In the coordinate system moving with the traveling wave, the streamfunction is given by
\begin{equation}
\phi(r)
=
\frac{p_{\xi}-1}{16}
\left(r^{2}-\alpha^{2}\right)^{2}
+
\frac{1}{4}
\left[
\sqrt{s}\,\gamma_{\xi}
-\frac{s}{2}\left(p_{\xi}-1\right)
\right]
\left[
2r^{2}\ln\left(\frac{r}{\alpha}\right)
+\alpha^{2}-r^{2}
\right]
-
\frac{c}{2}
\left(r^{2}-\alpha^{2}\right).
\end{equation}
As shown in \autoref{fig10}, the main effect of the electric field is to strengthen the recirculation within the wave crest, which is consistent with the results of \citet{Ding2014}. As $E_b$ increases, the region enclosed by the closed streamlines near the crest expands and the internal circulation becomes progressively stronger. The increase in the crest size is accompanied by an increase in the propagation speed, in agreement with the variation of $c$ shown in \autoref{fig9}($a$).To determine how the electric field influences Marangoni convection, \autoref{fig11} presents the total relative interfacial velocity $c_w$ at $Ma=0.1$, together with the velocity contribution $c_w^{(E_b)}$ obtained by retaining only the contribution of the electric stress gradient to $p_{\xi}$. As shown in \autoref{fig11}($a$), the entire distribution of $c_w$ shifts towards more negative values as $E_b$ increases, which indicates a substantial increase in the magnitude of the relative interfacial motion. However, this overall enhancement does not immediately lead to a corresponding increase in Marangoni convection. The formation of Marangoni convection depends on the nonuniform transport of surfactant along the interface, rather than only on the magnitude of $c_w$. More specifically, the electric field enhances the relative interfacial motion near the wave trough and therefore promotes the transport of surfactant towards the wave crest. At the same time, the circulation near the crest is also strengthened, which causes the surfactant accumulated in this region to be redistributed more rapidly. A competition therefore arises between the transport from the trough towards the crest and the redistribution near the crest. As $E_b$ continues to increase, the enhancement of the relative interfacial motion near the trough becomes more pronounced than that near the crest. The difference in transport efficiency between these two regions then becomes sufficiently large to strengthen the interfacial concentration gradient and, consequently, the Marangoni convection. This behaviour also shows that the RP mode and the Marangoni mode depend on the relative interfacial motion in different ways. The intensity of the RP mode is governed mainly by the overall magnitude of $c_w$, whereas the intensity of the Marangoni mode depends more strongly on the difference between the transport efficiencies near the crest and the trough. The electric field can therefore enhance the RP mode substantially even at smaller values of $E_b$, while Marangoni convection increases noticeably only after the spatial difference in surfactant transport becomes sufficiently large. This explains why $I_M$ changes only slightly at smaller values of $E_b$ but increases rapidly when $E_b$ becomes larger, as shown in \autoref{fig9}($d$). \autoref{fig11}($b$) further shows the velocity contribution $c_w^{(E_b)}$ obtained when only the electric stress gradient is retained. An important feature is that the velocity induced by the electric effect primarily transports fluid towards the wave crest, whereas its direct influence near the trough remains comparatively weak. This result indicates that the enhanced circulation near the crest originates mainly from the local forcing produced by the electric stress gradient. It also explains why the electric field increases both the crest size and the wave speed. In summary, the electric field acts through a physical mechanism that is distinct from Marangoni convection. The electric stress first strengthens the circulation near the wave crest and thereby increases the wave speed and the intensity of the relative interfacial motion. The electric field does not generate Marangoni convection directly. Instead, it changes the magnitude and spatial distribution of the relative interfacial velocity, which modifies surfactant transport between the crest and the trough and subsequently alters the interfacial concentration gradient and the Marangoni convection. This mechanism also accounts for the positive departure from linear superposition. The surfactant suppresses the RP mode by weakening the relative interfacial motion, whereas the electric stress acts directly to intensify it. The accompanying enhancement of Marangoni convection arises only as a consequence of the electrically modified interfacial motion and surfactant transport, and therefore cannot fully counteract the direct electrical amplification. As a result, the coupled response retains a stronger signature of the electric forcing, and the corresponding growth rate exceeds that predicted by linear superposition.

\section{Conclusion}
\label{sec:conclusion}
This study examines the coupled influence of an insoluble surfactant and a radial electric field on the stability of a viscous liquid film flowing down a vertical fiber. Starting from the governing equations in two dimensions, a reduced model in one dimension is derived using the long wave approximation to describe the evolution of the interface, surfactant transport, Marangoni stresses, and the normal electric stress. Linear stability analysis identifies two distinct disturbance modes: the Rayleigh-Plateau mode, which is associated primarily with interfacial deformation, and the Marangoni mode, which is associated primarily with disturbances in surfactant concentration. Increasing the Marangoni number suppresses the former, but destabilises the latter at sufficiently large values, leaving a finite parameter interval within which both modes remain stable at a prescribed wavenumber.

The influence of the radial electric field is determined by the position of the outer electrode $\beta$. When $\beta<e$, increasing the electric Weber number expands the unstable regions associated with both modes and narrows the stable interval separating them. The eventual connection of the two unstable regions occurs mainly because the region governed by the Rayleigh-Plateau instability extends towards larger Marangoni numbers. When $\beta>e$, the electric field has the opposite effect: both unstable regions contract as the electric Weber number increases, and the region associated with the Marangoni instability can be eliminated entirely. The combined use of a surfactant and an electric field with a stabilising effect may therefore provide a route to film stabilisation at a lower field strength.

The electric and Marangoni effects cannot, in general, be represented by simple linear superposition. Over finite ranges of wavenumber, the growth rate obtained when the two mechanisms act simultaneously exceeds the value predicted by adding their separate contributions. Continuation of the traveling wave solutions shows that increasing the electric Weber number generally increases the wave speed, the maximum interfacial radius, and the intensity of the relative interfacial motion. By contrast, Marangoni convection varies only weakly at smaller electric Weber numbers and becomes appreciably stronger only after the electric field reaches a sufficient intensity. Continuation in the Marangoni number further demonstrates that the essential mechanism through which the surfactant affects film stability is the competition between Marangoni convection and relative interfacial motion.

Analysis of the internal flow reveals the physical origin of this coupling. When $\beta<e$, the electric stress intensifies the recirculation beneath the wave crest and reshapes the spatial distribution of the relative interfacial motion. This redistribution modifies surfactant transport along the interface and reorganises the interfacial concentration field, thereby altering the Marangoni stress. The electric field therefore does not strengthen Marangoni convection directly, but instead modifies the competition between Marangoni convection and relative interfacial motion through its influence on the interfacial velocity field.

\begin{acknowledgments}
This research was supported by the National Natural Science Foundation of China (Grant No.12272026, U25A2032, U2341281, 12172317, 12502287, U25A20190), the State Key Laboratory of High-Efficiency Reusable Aerospace Transportation Technology (Grant No. RATT-ZZ04-2025033), the Beijing Natural Science Foundation (Grant No. L248008), the National Key Laboratory of Aerospace Liquid Propulsion Fund (Grant No. 2024JJ015003), and the Fundamental Research Funds for the Central Universities.
\end{acknowledgments}
\appendix
\section{mode selection}
\label{appA}
\begin{figure}
\centerline{\includegraphics[width=0.5\linewidth]{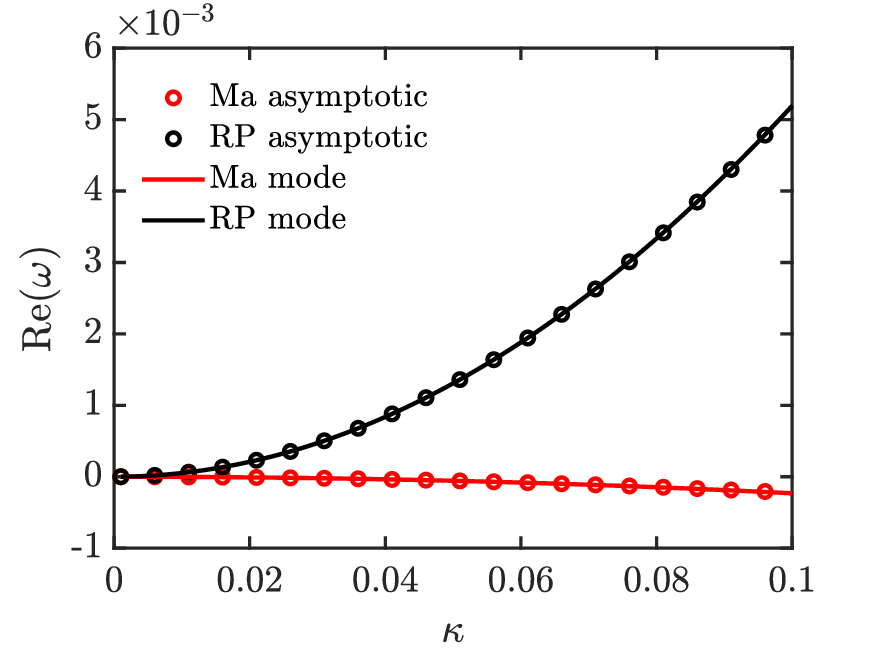}}
  \caption{Comparison between the asymptotic solutions in the long wave limit and the analytical solutions obtained over the full wavenumber range for the RP and Marangoni modes.}
  \label{fig12}
\end{figure}

The purpose of Appendix \ref{appA} is to present a procedure for identifying the two disturbance modes using the asymptotic solutions in the long wave limit. As shown in \autoref{fig2}($a$), the two eigenvalue branches intersect for certain values of $Ma$, whereas \eqref{fs: lambda} and \eqref{fs: A} provide only the larger and smaller eigenvalues according to their numerical values. These expressions therefore do not directly reveal the physical identity of each branch. To resolve this ambiguity, the solutions on either side of an intersection are first connected by requiring continuity of both the eigenvalues and their derivatives with respect to the wavenumber, thereby yielding two continuous branches. The propagation speed and growth rate of each branch in the long wave limit are then compared with the corresponding asymptotic solutions, which allows the branches to be identified as the RP mode and the Marangoni mode, respectively. In contrast to the treatment of \citet{Gao2026}, the axial curvature term is retained in the present asymptotic analysis so that the resulting approximation provides a more accurate description of the stability characteristics at finite wavenumbers. As shown in \autoref{fig12}, the asymptotic solutions agree closely with the analytical solutions obtained over the full wavenumber range in the long wave region, confirming the reliability of the proposed procedure for mode identification.

\section{Verification of numerical methods}\label{appB}
\begin{figure}
    \centering
    \begin{subfigure}[b]{0.45\textwidth}
        \centering
        \includegraphics[width=\textwidth, trim={0pt 0pt 0pt 0pt}, clip]{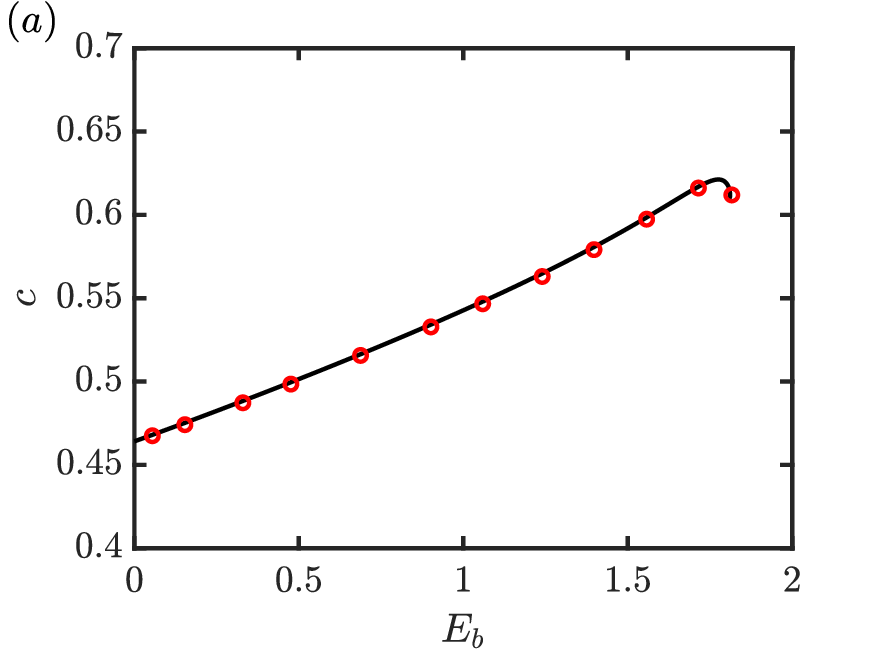}
    \end{subfigure}
    \hspace{0.02\textwidth}
    \begin{subfigure}[b]{0.45\textwidth}
        \centering
        \includegraphics[width=\textwidth, trim={0pt 0pt 0pt 0pt}, clip]{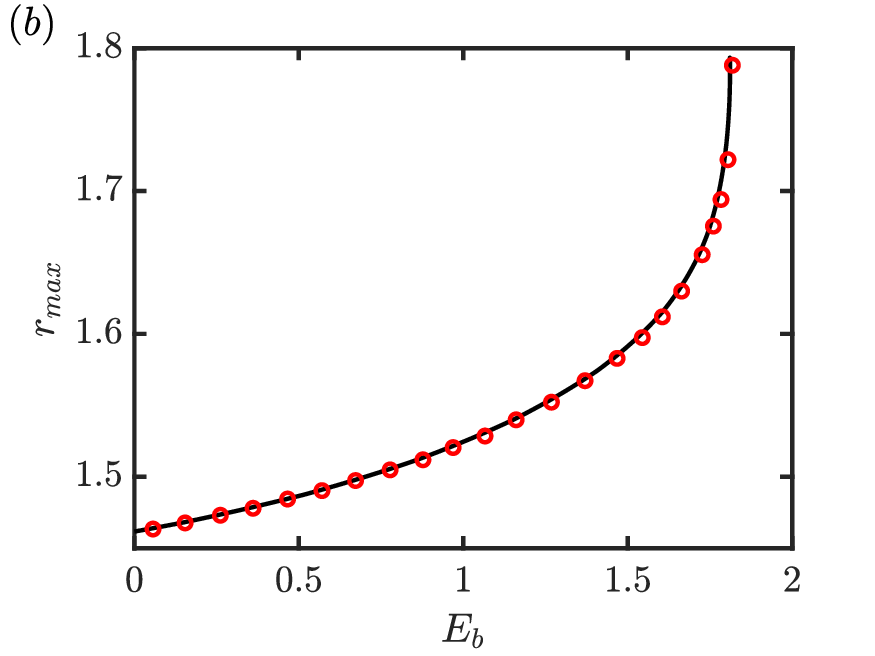}
    \end{subfigure}
    \caption{\raggedright Validation of the numerical implementation through comparison with the traveling wave solutions reported by \citet{Ding2014}, with $\alpha=0.28$, $\beta=\mathrm{e}^{0.9}$, $\varepsilon=0.23$, $L=1.64$, and $Ma=0$.} 
    \label{fig13}
\end{figure}

The purpose of this appendix is to validate the numerical method and its implementation in the present study. To this end, the results presented in figures~14($a,b$) of \citet{Ding2014} are adopted as a benchmark. The corresponding parameter values are $\alpha=0.28$, $\beta=\mathrm{e}^{0.9}$, $\varepsilon=0.23$, and $L=1.64$. Within the present formulation, the same problem is recovered by assigning these parameter values and setting $Ma=0$, which removes the contribution of the surfactant. As shown in \autoref{fig13}, the present numerical solutions agree closely with the reference results, thereby validating the numerical implementation and the procedure used to compute the traveling wave solutions.

\makeatletter
\renewcommand{\bibsection}{\section*{\refname}}
\makeatother

\bibliography{apssamp}%

@article{Nusselt1916,
  author  = {Nusselt, W.},
  title   = {Die Oberfl{\"a}chenkondensation des Wasserdampfes},
  journal = {Zeitschrift Vereines Deutsch Ing},
  volume  = {60},
  pages   = {541--546, 569--575},
  year    = {1916},
}

@article{Kapitza1948a,
  author  = {Kapitza, P. L.},
  title   = {Wave flow of thin layers of viscous liquid. Part I. Free flow},
  journal = {Zhurnal Eksperimentalnoi i Teoreticheskoi Fiziki},
  volume  = {18},
  pages   = {3--18},
  year    = {1948},
}

@article{Kapitza1948b,
  author  = {Kapitza, P. L.},
  title   = {Wave flow of thin layers of viscous liquids. Part II. Fluid flow in the presence of continuous gas flow and heat transfer},
  journal = {Zhurnal Eksperimentalnoi i Teoreticheskoi Fiziki},
  volume  = {18},
  pages   = {19--28},
  year    = {1948},
}

@article{Kapitza1949,
  author  = {Kapitza, P. L. and Kapitza, S. P.},
  title   = {Wave flow of thin layers of viscous liquids. Part III. Experimental research of a wave flow regime},
  journal = {Zhurnal Eksperimentalnoi i Teoreticheskoi Fiziki},
  volume  = {19},
  pages   = {105--120},
  year    = {1949},
}

@article{Benjamin1957,
  author  = {Benjamin, T. B.},
  title   = {Wave formation in laminar flow down an inclined plane},
  journal = {Journal of Fluid Mechanics},
  volume  = {2},
  number  = {6},
  pages   = {554--573},
  year    = {1957},
}

@article{Yih1963,
  author  = {Yih, C. S.},
  title   = {Stability of liquid flow down an inclined plane},
  journal = {The Physics of Fluids},
  volume  = {6},
  number  = {3},
  pages   = {321--334},
  year    = {1963},
}

@article{Benney1966,
  author  = {Benney, D.},
  title   = {Longwaves on liquid films},
  journal = {Journal of Mathematics and Physics},
  volume  = {45},
  number  = {1--4},
  pages   = {150--155},
  year    = {1966},
}

@article{Goren1962,
  author  = {Goren, S. L.},
  title   = {The instability of an annular thread of fluid},
  journal = {Journal of Fluid Mechanics},
  volume  = {12},
  number  = {2},
  pages   = {309--319},
  year    = {1962},
}

@article{Quere1990,
  author  = {Qu{\'e}r{\'e}, D.},
  title   = {Thin films flowing on vertical fibers},
  journal = {Europhysics Letters},
  volume  = {13},
  pages   = {721},
  year    = {1990},
}

@article{Frenkel1992,
  author  = {Frenkel, A.},
  title   = {Nonlinear theory of strongly undulating thin films flowing down vertical cylinders},
  journal = {Europhysics Letters},
  volume  = {18},
  pages   = {583},
  year    = {1992},
}

@article{Chang1999,
  author  = {Chang, H. C. and Demekhin, E. A.},
  title   = {Mechanism for drop formation on a coated vertical fibre},
  journal = {Journal of Fluid Mechanics},
  volume  = {380},
  pages   = {233--255},
  year    = {1999},
}

@article{Kalliadasis1994,
  author  = {Kalliadasis, S. and Chang, H. C.},
  title   = {Drop formation during coating of vertical fibres},
  journal = {Journal of Fluid Mechanics},
  volume  = {261},
  pages   = {135--168},
  year    = {1994},
}

@article{Camassa2014,
  author  = {Camassa, R. and Ogrosky, H. R. and Olander, J.},
  title   = {Viscous film flow coating the interior of a vertical tube. Part 1. Gravity-driven flow},
  journal = {Journal of Fluid Mechanics},
  volume  = {745},
  pages   = {682--715},
  year    = {2014},
}

@article{Camassa2016,
  author  = {Camassa, R. and Marzuola, J. L. and Ogrosky, H. R. and Vaughn, N.},
  title   = {Traveling waves for a model of gravity-driven film flows in cylindrical domains},
  journal = {Physica D: Nonlinear Phenomena},
  volume  = {333},
  pages   = {254--265},
  year    = {2016},
}

@Article{Kliakhandler2001,
 author      = {I.L. Kliakhandler and S.H. Davis and  S.G. Bankoff},
 title       = {Viscous beads on vertical fibre},
 journal     = {J. Fluid Mech.},
 year        = {2001},
 volume      = {429},
 pages       = {381-390}
 }

@Article{Craster2006,
  author  = {R. V. Craster and O. K. Matar},
  title   = {On viscous beads flowing down a vertical fibre},
  journal = {J. Fluid Mech.},
  year    = {2006},
  volume  = {553},
  pages   = {85--105}
}

@Article{Duprat2007,
 author      = {C. Duprat and C.R. Quil and S. Kalliadasis and F.G. Dauphiné},
 title       = {Absolute and convective Instabilities of a Viscous Film Flowing Down a Vertical Fiber},
 journal     = {Phys. Rev. Lett.},
 year        = {2007},
 volume      = {98},
 pages       = {244502 }
 }

@Article{Ding2019,
  author  = {Z. J. Ding and A. P. Willis},
  title   = {Relative periodic solutions in conducting liquid films flowing down vertical fibres},
  journal = {J. Fluid Mech.},
  year    = {2019},
  volume  = {873},
  pages   = {835--855}
}

@Article{RuyerQuil2008,
  author  = {C. Ruyer-Quil and P. Trevelyan and C. Duprat and F. G. Giorgiutti-Dauphin{\'e} and S. Kalliadasis},
  title   = {Modelling film flows down a fibre},
  journal = {J. Fluid Mech.},
  year    = {2008},
  volume  = {603},
  pages   = {431--462}
}

@Article{Kishore2012,
  author  = {V. A. Kishore and D. Bandyopadhyay},
  title   = {Electric field induced patterning of thin coating on fiber surfaces},
  journal = {J. Phys. Chem. C},
  year    = {2012},
  volume  = {116},
  pages   = {6215--6221}
}

@Article{Li2009,
  author  = {B. Li and Y. Li and G. K. Xu and X. Q. Feng},
  title   = {Electric field induced patterning of thin coatings on fiber surfaces},
  journal = {J. Phys.-Condens. Mat.},
  year    = {2009},
  volume  = {21},
  pages   = {445006}
}

@article{Boulogne2012Instability,
  author  = {Boulogne, F. and Pauchard, L. and Giorgiutti-Dauphin{\'e}, F.},
  title   = {Instability and morphology of polymer solutions coating a fibre},
  journal = {Journal of Fluid Mechanics},
  volume  = {704},
  pages   = {232--250},
  year    = {2012},
}

@article{Gabbard2021Asymmetric,
  author  = {Gabbard, C. T. and Bostwick, J. B.},
  title   = {Asymmetric instability in thin-film flow down a fiber},
  journal = {Physical Review Fluids},
  volume  = {6},
  pages   = {034005},
  year    = {2021},
}

@article{Xu2017FoodOilSurfaceTension,
  author  = {Xu, Tong and Rodriguez-Martinez, Veronica and Sahasrabudhe, Shreya N. and Farkas, Brian E. and Dungan, Stephanie R.},
  title   = {Effects of Temperature, Time and Composition on Food Oil Surface Tension},
  journal = {Food Biophysics},
  volume  = {12},
  pages   = {88--96},
  year    = {2017},
}

@article{Sawada2005FluoroalkylCooligomers,
  author  = {Sawada, Hideo and Horiuchi, Hitomi and Kawase, Tokuzo and Oharu, Kazuya and Nakagawa, Hideki and Kaneda, Isamu},
  title   = {Synthesis and Applications of Silicone Oil--Soluble Fluoroalkyl End-Capped Cooligomers},
  journal = {Journal of Applied Polymer Science},
  volume  = {96},
  number  = {4},
  pages   = {1467--1476},
  year    = {2005},
}

@PhdThesis{Kas-Danouche2002,
  author = {S.A. Kas-Danouche},
  title = {Nonlinear interfacial stability of core-annular film flows in the presence of surfactants},
  school = {New Jersey Institute of Technology and Rutgers University},
  year = {1992},
  address = {}
}

@Article{Li2023,
 author      = {S. Li and Y.Z. Chen and Z. Cheng and J. Peng},
 title       = {The role of soluble surfactant in the linear instability of a film coating inside a tube},
 journal     = {J. Fluid Mech.},
 year        = {2023},
 volume      = {973},
 pages       = {A46}
 }

@Article{Pereira2008,
 author      = {A. Pereira and S. Kalliadasis},
 title       = {Dynamics of a falling film with solutal Marangoni effect},
 journal     = {Phys. Rev. E},
 year        = {2008},
 volume      = {78},
 pages       = {036312}
 }

@article{Gao2026,
  author  = {Gao, Jun and Yang, Xiaocong and Zhu, Senlin and Fu, Qingfei and Yang, Lijun},
  title   = {Dynamics of viscous beads on vertical fibers with insoluble surfactants},
  journal = {Physical Review Fluids},
  volume  = {11},
  pages   = {053901},
  year    = {2026},
}

@article{Wong1996,
  author  = {Wong, Harris and Rumschitzki, David and Maldarelli, Charles},
  title   = {On the surfactant mass balance at a deforming fluid interface},
  journal = {Physics of Fluids},
  volume  = {8},
  number  = {11},
  pages   = {3203--3204},
  year    = {1996},
}

@article{Stone1990,
  author  = {Stone, H. A.},
  title   = {A simple derivation of the time-dependent convective-diffusion equation for surfactant transport along a deforming interface},
  journal = {Physics of Fluids A: Fluid Dynamics},
  volume  = {2},
  number  = {1},
  pages   = {111--112},
  year    = {1990},
}

@article{Halpern2003,
  author  = {Halpern, D. and Frenkel, A. L.},
  title   = {Destabilization of a creeping flow by interfacial surfactant: Linear theory extended to all wavenumbers},
  journal = {Journal of Fluid Mechanics},
  volume  = {485},
  pages   = {191--220},
  year    = {2003},
}

@Article{Blyth2006,
  author  = {M. G. Blyth and H. Luo and C. Pozrikidis},
  title   = {Stability of axisymmetric core--annular flow in the presence of an insoluble surfactant},
  journal = {J. Fluid Mech.},
  year    = {2006},
  volume  = {548},
  pages   = {207--235}
}

@Article{Blyth2004,
  author  = {M. G. Blyth and C. Pozrikidis},
  title   = {Effect of surfactant on the stability of film flow down an inclined plane},
  journal = {J. Fluid Mech.},
  year    = {2004},
  volume  = {521},
  pages   = {241--250}
}

@Article{Nair2020,
  author  = {A. Nair and G. Sharma},
  title   = {Stability of surfactant-laden liquid film flow over a cylindrical rod},
  journal = {Phys. Rev. E},
  year    = {2020},
  volume  = {102},
  pages   = {023111}
}

@Article{Wei2005,
  author  = {H. H. Wei and D. Rumschitzki},
  title   = {The effects of insoluble surfactants on the linear stability of a core--annular flow},
  journal = {J. Fluid Mech.},
  year    = {2005},
  volume  = {541},
  pages   = {115--142}
}

@Article{Wray2012,
  author  = {A. W. Wray and O. Matar and D. T. Papageorgiou},
  title   = {Non-linear waves in electrified viscous film flow down a vertical cylinder},
  journal = {IMA J. Appl. Math.},
  year    = {2012},
  volume  = {77},
  pages   = {430--440}
}

@Article{Wray2013,
  author  = {A. W. Wray and D. T. Papageorgiou and O. Matar},
  title   = {Electrified coating flows on vertical fibres: enhancement or suppression of interfacial dynamics},
  journal = {J. Fluid Mech.},
  year    = {2013},
  volume  = {735},
  pages   = {427--456}
}

@Article{Ding2014,
 author      = {Z.J. Ding and J.L. Xie and T.N. Wong and R. Liu},
 title       = {Dynamics of liquid films on vertical fibres in a radial electric field},
 journal     = {J. Fluid Mech.},
 year        = {2014},
 volume      = {752},
 pages       = {66-89}
 }

@article{Yang_2025,
  title={Nonlinear dynamics analysis of an ionic surfactant-laden thread in a radial electric field},
  volume={1025},
  journal={Journal of Fluid Mechanics},
  author={Yang, Xiaocong and Gao, Jun and Sun, Hu and Liu, Qiyou and Qiao, Wentong and Ji, Bingqiang and He, Dongdong and Fu, Qingfei},
  year={2025},
  pages={A36},
}

\end{document}